\documentclass[acmsmall,screen]{acmart}
\usepackage{amsmath}
\usepackage{caption}
\usepackage{subfigure}
\usepackage{adjustbox}
\usepackage{multirow}
\usepackage{listings}

\definecolor{codegreen}{rgb}{0,0.6,0}
\definecolor{codegray}{rgb}{0.5,0.5,0.5}
\definecolor{codepurple}{rgb}{0.58,0,0.82}
\definecolor{backcolour}{rgb}{0.95,0.95,0.92}

\lstdefinestyle{mystyle}{
  backgroundcolor=\color{backcolour}, 
  commentstyle=\color{codegreen},
  keywordstyle=\color{magenta},
  numberstyle=\tiny\color{codegray},
  stringstyle=\color{codepurple},
  basicstyle=\ttfamily\footnotesize,
  breakatwhitespace=false,    
  language=Python,
  breaklines=true,                 
  captionpos=b,                    
  keepspaces=true,                 
  numbers=left,                    
  numbersep=5pt,                  
  showspaces=false,                
  showstringspaces=false,
  showtabs=false,                  
  tabsize=2,
  float=tp
}
\usepackage{threeparttable}

\AtBeginDocument{
  }

\setcopyright{cc}
\setcctype{by-sa}
\copyrightyear{2026}
\acmYear{2026}
\acmDOI{10.1145/3777387}

\acmJournal{TRETS}
\acmVolume{19}
\acmNumber{1}
\acmArticle{13}
\acmMonth{3}

\begin{document}

\title{da4ml: Distributed Arithmetic for Real-time Neural Networks on FPGAs}

\author{Chang Sun}

\email{chsun@cern.ch}
\orcid{0000-0003-2774-175X}
\affiliation{
  \institution{California Institute of Technology}
  \city{Pasadena}
  \state{CA}
  \country{USA}
}

\author{Zhiqiang Que}
\email{z.que@imperial.ac.uk}
\orcid{0000-0002-9263-6529}
\affiliation{
  \institution{Imperial College London}
  \city{London}
  \country{UK}
}

\author{Vladimir Loncar}
\email{vloncar@cern.ch}
\orcid{0000-0003-3651-0232}
\affiliation{
  \institution{CERN}
  \city{Geneva}
  \country{Switzerland}
}
\affiliation{
  \institution{Institute of Physics Belgrade}
  \city{Belgrade}
  \country{Serbia}
}

\author{Wayne Luk}
\email{w.luk@imperial.ac.uk}
\orcid{0000-0002-6750-927X}
\affiliation{
  \institution{Imperial College London}
  \city{London}
  \country{UK}
}

\author{Maria Spiropulu}
\email{smaria@caltech.edu}
\orcid{0000-0001-8172-7081}
\affiliation{
  \institution{California Institute of Technology}
  \city{Pasadena}
  \state{CA}
  \country{USA}
}

\begin{abstract}
  Neural networks with a latency requirement on the order of microseconds, like the ones used at the CERN Large Hadron Collider, are typically deployed on FPGAs fully unrolled and pipelined. A bottleneck for the deployment of such neural networks is area utilization, which is directly related to the required constant matrix-vector multiplication (CMVM) operations. In this work, we propose an efficient algorithm for implementing CMVM operations with distributed arithmetic on FPGAs that simultaneously optimizes for area consumption and latency. The algorithm achieves resource reduction similar to state-of-the-art algorithms while being significantly faster to compute. The proposed algorithm is open-sourced and integrated into the \texttt{hls4ml} library, a free and open-source library for running real-time neural network inference on FPGAs. We show that the proposed algorithm can reduce on-chip resources by up to a third for realistic, highly quantized neural networks while simultaneously reducing latency, enabling the implementation of previously infeasible networks.
\end{abstract}



\maketitle

\makeatletter
\newcommand{\ACM@publishedfoot}{\footnotesize Published in \@journalNameShort,
  Vol.~\@acmVolume, No.~\@acmNumber, Article~\@acmArticle.}
\fancypagestyle{standardpagestyle}{
  \fancyhf{}
  \renewcommand{\headrulewidth}{\z@}
  \renewcommand{\footrulewidth}{\z@}
  \fancyhead[LE]{\@headfootfont\thepage}
  \fancyhead[RO]{\@headfootfont\thepage}
  \fancyhead[RE]{\@headfootfont\@shortauthors}
  \fancyhead[LO]{\@headfootfont\shorttitle}
  \fancyfoot[RO,LE]{\ACM@publishedfoot}
}
\fancypagestyle{firstpagestyle}{
  \fancyhf{}
  \renewcommand{\headrulewidth}{\z@}
  \renewcommand{\footrulewidth}{\z@}
  \fancyfoot[RO,LE]{\ACM@publishedfoot}
}
\pagestyle{standardpagestyle}
\thispagestyle{firstpagestyle}
\makeatother

\section{Introduction}

Due to the need for low latency and high throughput in applications, edge computing has significantly increased its importance for real-time neural network inference~\cite{edge-survey}. While typical real-time inference applications have latency constraints of a few milliseconds~\cite{realtime1,realtime2,realtime3}, certain specific applications require sub-microsecond inference latency. For instance, at the CERN Large Hadron Collider (LHC)~\cite{lhc1995large}, hundreds of terabytes of data are generated by the detectors every second from proton-proton collisions at a rate of 40 MHz. It is technically infeasible to store all collision events. This enormous data throughput is reduced by a hardware system, $trigger$, which filters the data in real-time at the same rate. It determines which event should be kept for offline processing or discarded, and the final decision must be made within a few microseconds~\cite{cms-tdr-021,atlas-tdr-029}. The trigger's accuracy is vital to keep only the relevant events for physics studies, effectively managing the downstream data rate by reducing it by two orders of magnitude. Over one thousand of field-programmable gate arrays (FPGAs) are currently used in the trigger system, where several algorithms run in parallel on each FPGA. As a result, resources are scarce, and the footprint of each algorithm must be minimal. In anticipation of the LHC's upgrade to the High-Luminosity LHC (HL-LHC)~\cite{hl-lhc}, Machine Learning (ML) techniques are being actively explored to enhance the current trigger system~\cite{cms-tdr-021,atlas-tdr-029,axol1tl}. However, integrating demanding models under such strict resource and latency constraints remains challenging.

To meet the latency requirements, the neural networks used in such applications are typically deployed on FPGAs fully unrolled and pipelined (usually with an initiation interval of one). \texttt{hls4ml}, a free and open-source library for generating deployable hardware designs for neural network inferences on FPGAs, is widely used for implementing such neural networks on FPGAs~\cite{atlas-tdr-029,cms-tdr-021,tgc,axol1tl,hls4ml,adpaper}. The library parses the architectures of the neural networks and generates the corresponding high-level synthesis (HLS) projects for FPGAs with a set of fixed templates for common neural network layers or operations. The generated HLS projects are then synthesized into hardware description language (HDL) code by the corresponding HLS backends.

Frequently, the most resource-consuming part of the designs is the constant matrix-vector multiplication (CMVM) operations in the neural networks, such as those in the dense or convolutional layers. As the final design is fully unrolled, the CMVM operations are typically implemented primarily with distributed arithmetic (DA)~\cite{fir2} by the HLS backends, as suggested by the high lookup-table (LUT) utilization with moderate to low digital signal processor (DSP) utilization in the synthesized designs.

In this work, we present \texttt{da4ml}, a fast, scalable, and accurate optimization framework for CMVM targeting ultra-low-latency neural networks on FPGAs.
Most existing exact multiplierless CMVM algorithms are either too slow (e.g., $H_{cmvm}$: $\mathcal{O}(N^3)$\footnote{$N$ is the number of non-zero signed digits needed to represent the constant matrix.} time~\cite{hcmvm}, hours for moderate-size matrices), or miss significant opportunities for optimization (e.g., SCMVM: $\mathcal{O}(N^2)$ time~\cite{scmvm}, but cannot capture differently scaled subexpressions).
The proposed \texttt{da4ml} combines a novel graph-based decomposition with cost-aware Common Subexpression Elimination (CSE), which
retains full numerical precision (not approximate), captures subexpression reuse even across differently scaled terms and signed digits, and has a much better asymptotic complexity $\mathcal{O}(N^2)$ and a runtime that is five orders of magnitude faster (Table~\ref{tab:random_mat}).
This makes it practical for realistic, large-scale networks in ultra-low-latency environments such as the CERN LHC.

To the best of our knowledge, this is the first open-source, end-to-end DA-based neural network compiler with optimized CMVM operations. By tightly integrating our framework with \texttt{hls4ml}, it provides a drop-in complement for the default CMVM implementation in \texttt{hls4ml}. This tight integration into an established toolchain (the \texttt{hls4ml} workflow) lowers the barrier to adoption for the broader High Energy Physics (HEP) and FPGA communities. The framework also supports direct RTL generation without going through HLS for fast prototyping and easier integration into existing RTL workflows. The practical impact of the proposed framework has been demonstrated by enabling the production deployment of the AXOL1TL~\cite{axol1tl} anomaly detection trigger at the CMS experiment, where it significantly reduces the resource utilization and timing closure of the synthesized design.

The contributions of this work are as follows:
\begin{itemize}
  \item We propose \texttt{da4ml}, a novel, performant, and scalable CMVM optimization framework with a hybrid algorithm that combines graph-based decomposition with cost-aware CSE. It offers orders of magnitude faster runtime than the prior state of the art while achieving comparable resource efficiency.
  \item We implement the proposed \texttt{da4ml} framework as an open-source library\footnote{\url{https://github.com/calad0i/da4ml}} and tightly integrate it into the widely used \texttt{hls4ml} tool as a drop-in solution for FPGA-based ML designs. \texttt{da4ml} can also work standalone and generate synthesizable RTL directly, allowing users to bypass \texttt{hls4ml} or HLS where needed. In addition, \texttt{da4ml} has enabled production deployment of the AXOL1TL~\cite{axol1tl} anomaly detection trigger at CMS with improved resource efficiency and timing closure.
  \item We evaluate the proposed framework on both synthetic CMVM benchmarks and realistic neural networks for the CERN LHC trigger system. The results show significant improvements in hardware resource utilization, design latency, and compilation speed, demonstrating the framework's practicality for ultra-low-latency FPGA applications.

\end{itemize}

We organize the rest of the paper as follows. Section~\ref{sec:related_work} reviews the related work on CMVM algorithms and ultra-low latency neural networks on FPGAs. Section~\ref{sec:problem_formulation} formalizes the CMVM problem and the optimization objectives. Section~\ref{sec:da4ml_algorithm} details the proposed \texttt{da4ml} algorithm. Section~\ref{sec:impl} describes the implementation of the proposed framework and its integration into the ecosystem. Section~\ref{sec:experiments} evaluates the proposed framework on both synthetic benchmarks and realistic neural networks. Finally, we conclude the paper in Section~\ref{sec:conclusion}.

\newcommand{\hcmvm}{$\mathrm{H}_\mathrm{cmvm}$}
\section{Background and Related Work}\label{sec:related_work}

\subsection{Constant Matrix-Vector Multiplication}

Constant matrix-vector multiplication (CMVM) is a common problem in digital signal processing (DSP) applications. The problem involves computing an operation of the form $\vec y^\mathrm{T}=\vec x^\mathrm{T}M$, where $M$ is a constant integer matrix and $\vec x$ is a vector of input values. The problem is well studied in the literature, and there are multiple algorithms proposed for solving it.

Distributed arithmetic (DA) is a multiplierless method for implementing the multiplication-accumulation (MAC) operations in hardware by replacing them by shift-and-add (or subtract) operations, which are usually mapped to LUTs on FPGAs. The method is particularly useful for applications where extremely low latency and high throughput are required.

The DA problem has been studied in the literature for a long time, and multiple algorithms have been proposed for solving it. Many of the existing works focus on the multiple constant multiplication (MCM) problem, which is a special case of the CMVM problem where the variable vector $\vec x$ is of size one, commonly used in finite impulse response (FIR) filters~\cite{hcmvm}, such as~\cite{mmcm,da-fir,fir2,mmcm2,fir,cse-mcm}.

The works ~\cite{hcmvm,scmvm,mcmt,rag-cmvm,pcmvm,mod-cmvm,most-common-cmvm} propose algorithms for the CMVM problem, which is a more general case of the MCM problem. \cite{most-common-cmvm} applies CSE to the CMVM problem by recursively removing the most common two-term subexpression from the problem. \cite{mod-cmvm} uses a similar approach but also takes into account the conflict between the subexpressions. In contrast, \cite{rag-cmvm} uses a graph based approach, which transforms subgraphs in the constructed adder tree without increasing the maximum adder depth, the longest path counted by the number of adders from the input to the output that can be used as a proxy for latency for the final design.

Though dated more than a decade, to the best of our knowledge, the state-of-the-art algorithm for implementing a CMVM operation with an adder tree without precision loss is still the \hcmvm algorithm proposed in $H_\mathrm{CMVM}$~\cite{hcmvm}. The algorithm aggressively searches for possible transformations of the CMVM problem into potentially simpler subproblems and evaluates the cost of each using a heuristic similar to \cite{mod-cmvm}. The algorithm also supports specifying the maximum allowed adder depth for the generated adder tree by limiting the search phase space. However, this algorithm is computationally expensive, which requires an order of 1000 CPU-seconds to optimize a random $16\times 16$, 8-bit matrix. Moreover, the runtime scaling with respect to $N$, the number of non-zero signed digits required to represent the constant matrix, lies between $\mathcal{O}(N^3)$ and $\mathcal{O}(N^{3.5})$. This makes it impractical for optimizing any moderately large matrices used in neural networks. In contrast, we propose a new algorithm for optimizing the CMVM problem that achieves resource efficiency comparable to~\cite{hcmvm} while significantly reducing computational cost, making it practical for larger matrices.

Scalable CMM~\cite{scmvm} proposes a scalable algorithm for the CMVM problem, which is demonstrated to be runtime-efficient and practically useful for up to $100\times 100$ constant matrices. Unlike previously mentioned CSE-based approaches using two-term subexpressions, the algorithm considers multi-term common subexpressions with greedy CSE for efficient processing. However, this algorithm fails to capture common subexpressions with different power-of-two scaling factors, and it does not account for possible negative values in the weights. MCMT~\cite{mcmt} adopts a similar multi-term common subexpression approach but uses simulated annealing to optimize the cost of the adder tree for better resource utilization at the expense of algorithm runtime. While shifted common subexpressions are supported, the algorithm can only identify them in a coarse-grained fashion (i.e., a uniform shift across the whole row or column of the constant matrix). Signed digit representation is stated to be supported, but no discrete method is given, and its impact on identifying multi-term common subexpressions is not discussed. To avoid these drawbacks, we design our algorithm to effectively capture common subexpressions with different shifts and signs while preserving runtime efficiency, enabling more efficient optimization for a broader class of constant matrices.

For neural networks, implementing CMVM operations without precision loss is not always required. Ref.~\cite{cmvm-cc} uses computation coding for approximate CMVM operations by decomposing the constant matrix into the product of multiple matrices containing only powers-of-two. By applying this method to embedded neural networks, the method demonstrates minimal accuracy loss while improving resource utilization by 3 to 6-fold. Similarly, AURA~\cite{aura} proposes another method based on 0-1 linear programming to optimize the CMVM problem approximately. However, for already highly quantized neural networks, such as those trained with HGQ~\cite{hgq} or QKeras~\cite{qkeras} with low bitwidths, the weights usually carry non-negotiable gradients on the losses, and such additional approximations in the implementation stage are not preferred. Hence, This work focuses on the implementation of CMVM operations without any approximations, preserving full numerical precision throughout.

\subsection{Ultra-Low Latency Neural Networks}
Neural networks with a latency requirement on the order of $\mathcal{O}(1)$ $\mu$s are required for the trigger systems of the Large Hadron Collider due to size limitation of the on-detector buffers. Simultaneously, the networks are required to produce one inference at least every 25 ns due to the collision rate. To satisfy these constraints, these neural networks are typically heavily quantized and pruned and deployed on FPGAs fully unrolled with an initiation interval (II) of one~\cite{atlas-tdr-029,cms-tdr-021,tgc,axol1tl,hls4ml,adpaper}.

As the networks are fully unrolled, on-chip resource consumption is of major concern. In our experience, the most resource-consuming part of the designs is usually the matrix-vector multiplications in the fully connected or convolutional layers. The standard approach for optimizing resource utilization is to quantize and prune neural networks, reducing the model size and computational cost, and hence the FPGA resource usage. In this work, we go further by introducing an efficient algorithm for optimizing CMVM operations based on distributed arithmetic, which reduces resource utilization by up to one-third and improves timing closure, enabling previously infeasible designs to meet the stringent LHC trigger system constraints.

\subsection{\texttt{hls4ml}}

\texttt{hls4ml}~\cite{hls4ml} is a template library for translating quantized neural networks into HLS projects for FPGAs. \texttt{hls4ml} provides a high-level interface for users to interface popular machine learning frameworks with HLS backends, and enables users to deploy neural networks on FPGAs with minimal effort.

This library is widely used for real-time inference applications. Notably, it has been deployed at the Compact Muon Solenoid (CMS) experiment in its L1 trigger for anomaly detection~\cite{axol1tl} and is being studied for the upgrade of the trigger system at the ATLAS experiment~\cite{tgc}.

The library has a large built-in template library for common neural network layers, each customized and optimized for different HLS backends. In this work, we use \texttt{hls4ml} with the Vivado~\cite{vivado} and Vitis HLS~\cite{vitis} backends to implement the proposed algorithm, and we use its implementation as a baseline for comparison.

\section{Problem Formulation}
\label{sec:problem_formulation}
The objective is to implement the CMVM operation, $\vec y^\textrm{T}=\vec x^\mathrm{T}M$, as an adder tree with minimum LUTs on FPGAs under a delay constraint. The delay constraint (DC) is defined by the maximum of additional adder depth with respect to the minimal adder depth possible. The target application of this work is the real-time inference of quantized neural networks on FPGAs targeting the LHC trigger system, with possible extension to other general DSP applications. For the target application, the neural networks are required to have an initiation interval $\le 25$ ns. For this purpose, when implementing the CMVM operations with adder trees, we expect multiple adder depths within one pipeline stage. Hence, though the final design is expected to be pipelined, we expect the number of LUTs used for implementing the adders to dominate the resource utilization with moderate to low register usage.

In the adder tree, the dominant operation is in the form of $a \pm (b << s)$, where $a$ and $b$ are the variable inputs to the adder. The bit-shift $s$ and the sign for the input $b$ are known at compilation time. The expected cost of this operation is formulated as the number of full and half adders required to implement the operation (i.e., the number of output bits conditioned on more than one input). We denote the bitwidths for $a$ and $b$ as $bw_{a}$ and $bw_{b}$, respectively. When there is at least one bit overlapping between the two operands (i.e., $\max(bw_{a}, bw_{b}) > s$), the simplified expected cost is given by

\begin{equation}
  \mathrm{cost}(bw_{a}, bw_{b}, s, \mathrm{sign}) = \max(bw_{a}, bw_{b} + s) - \min(0, s) + 1. \label{eq:cost_adder}
\end{equation}

In practice, the logic delay of each adder with different input bitwidths will be different. However, as we notice that the majority of the delay in the adder trees is due to routing, we assume each adder to have the same delay and model the overall delay approximately by the adder depth for simplicity, following~\cite{hcmvm, aura, rag-cmvm}.

\section{\texttt{da4ml} Algorithm}
\label{sec:da4ml_algorithm}

\subsection{Notations}
For a number of fixed-point type, it takes the general form of \texttt{fixed<$S$, $W$, $I$>}, where $S$ is the sign bit, $W$ is the total bitwidth, and $I$ is the number of integer bits including the sign bit. In the algorithm, we denote the fixed-point numbers by \textit{quantized interval}, namely, by their low value, high value, and the step size, $[l,h,\delta]$. For a generic fixed-point number in the form of \texttt{fixed<$S$, $W$, $I$>}, we have $l=-S \cdot 2^{I-S}$, $h=2^{I-S}-2^{-W+I}$, and $\delta=2^{-W+I}$. This notation is useful to track the required bitwidths when accumulating a large number of fixed-point numbers, as otherwise one carry bit will always need to be added to prevent overflow.

The notation we use in the algorithm is summarized in Table~\ref{tab:parameters}. In particular, $M[d_{in},d_{out}]$, $qint_{in}[d_{in}]$, $depth_{int}[d_{in}]$, and $dc$ are the input parameters to the algorithm for a single CMVM operation.

\begin{table*}[htbp]
  \centering
  \caption{List of parameters used in the proposed \texttt{da4ml} algorithm}
  \label{tab:parameters}
  \begin{tabular}{l p{0.7\textwidth}}
    \toprule
    \textbf{Parameter}            & \textbf{Description}                                                                                                        \\
    \midrule
    $d_{in}$                      & The number of input elements.                                                                                               \\
    $d_{out}$                     & The number of output elements.                                                                                              \\
    $M[d_{in},d_{out}]$           & The input constant matrix containing fixed-point numbers.                                                                   \\
    $dc$                          & Delay constraint, maximum number of extra adder depth permitted beyond $\mathrm{depth}_{min}=\lceil \log_2(d_{in}) \rceil$. \\
    $qint_{in}[d_{in}]$           & An array representing the quantized intervals of the input vector.                                                          \\
    $l, h, \delta$                & The low, high, and step size that uniquely defines a quantized interval.                                                    \\
    $depth_{int}[d_{in}]$         & The adder depth associated with each element of the input vector.                                                           \\
    $W, I, S$                     & The width, integer bits (include sign bit if present), and sign bit of a fixed-point number.                                \\
    $s$                           & The bit-shift applied to an operand in a two-term subexpression.                                                            \\
    $bw_M$                        & The bitwidth of the constant matrix $M$.                                                                                    \\
    \midrule
    \multicolumn{2}{l}{\textbf{First Stage}}                                                                                                                    \\
    $\vec{v}_i$                   & A vector corresponding to the i-th column of the constant matrix $M$.                                                       \\
    $v_i$                         & A vertex corresponds to the column vector $\vec{v}_i$.                                                                      \\
    $\vec{v}_0$                   & The root vertex in the graph, associated with the zero vector ($\vec{0}$).                                                  \\
    $M_1, M_2$                    & The two submatrices such that $M = M_1M_2$.                                                                                 \\
    \midrule
    \multicolumn{2}{l}{\textbf{Second Stage}}                                                                                                                   \\
    $M$                           & Overloaded to the input matrix for the second stage, either $M_1$ or $M_2$.                                                 \\
    $\hat{M}$                     & The normalized version of matrix $M$.                                                                                       \\
    $M_{expr}[d_{in},d_{out}, B]$ & The CSD representation of the matrix $\hat{M}$, with values in $\{-1, 0, 1\}$.                                              \\
    $B$                           & The span of powers of the CSD digits.                                                                                       \\
    $L_{impl}$                    & A list containing the implemented values.                                                                                   \\
    $a, b$                        & The two input values for a two-term subexpression.                                                                          \\
    \bottomrule
  \end{tabular}
\end{table*}

\begin{figure}
  \centering
  \includegraphics[width=1.0\textwidth]{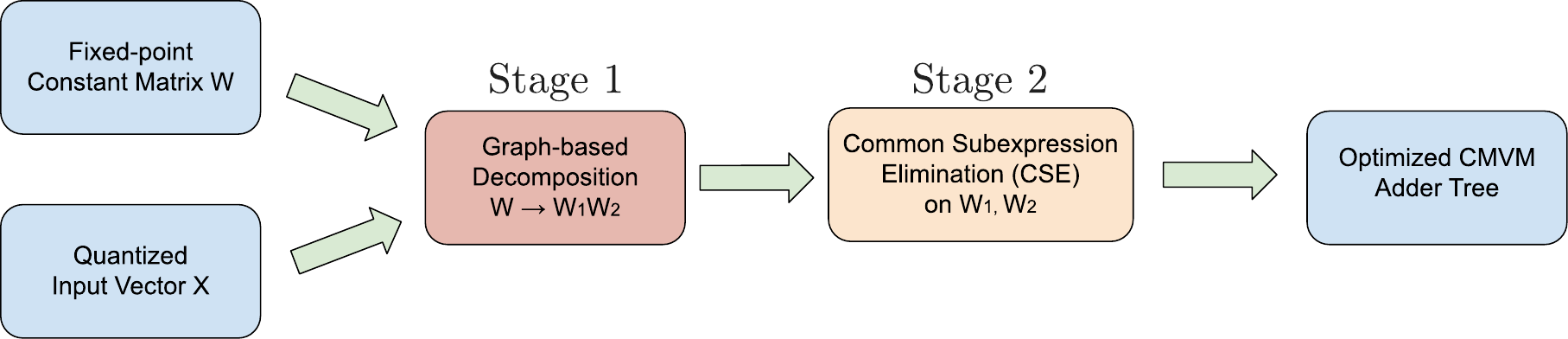}
  \caption{Overview of the proposed \texttt{da4ml} automatic optimization flow for CMVM on FPGAs. The algorithm first decomposes the constant matrix $M$ into two submatrices $M_1$ and $M_2$ using a graph-based approach that captures shared structure across columns. It then applies CSE on both submatrices to minimize redundant computations. The resulting optimized adder tree significantly reduces LUT usage and the latency.}
  \label{fig:approach}
  \Description{A flowchart showing the two-stage \texttt{da4ml} optimization process. The inputs, a constant matrix W and an input vector X, are first processed by Stage 1, labeled "Graph-based Decomposition". The output of Stage 1 feeds into Stage 2, labeled "Common Subexpression Elimination". The final output of the flow is an "Optimized CMVM Adder Tree", representing the hardware implementation.}
\end{figure}

\subsection{Overview}

The \texttt{da4ml} is a CSE-based hybrid algorithm. Like other CSE-based algorithms, the algorithm operates on a discrete representation of the constant matrix. In this work, following~\cite{hcmvm, aura, mod-cmvm,most-common-cmvm}, we adopt the canonical signed digit (CSD)~\cite{csd} representation. CSD is a signed digit representation of a number that never has two consecutive non-zero digits, and the number of non-zero digits is guaranteed to be minimal. Hence, for a number with $x$ digits, the CSD representation has at most $\lfloor x/2 + 1\rfloor$ non-zero digits, which is $\sim 1/3$ of the total number of digits on average.

As $M$ contains only fixed-point integers, the algorithm first normalizes it by applying bit-shifts across the rows and columns such that no row/column has all entries even except for zeros. The resultant scaling factors are recorded and will be applied to the input/output vectors.

The proposed \texttt{da4ml} algorithm first decomposes the constant matrix $M$ into two submatrices $M_1$ and $M_2$ using a graph-based approach that captures high-level common pattern across rows. It then applies CSE on both submatrices to minimize redundant computations, as shown in Figure~\ref{fig:approach}. The resulting optimized adder tree significantly reduces LUT usage and the latency.

\subsection{First stage: Graph-based Decomposition}

In the first stage, the algorithm decomposes the constant matrix into two submatrices, $M_1$ and $M_2$ such that $M=M_1M_2$. Inspired by \hcmvm, we propose this stage to exploit the high-level similarity between the different columns of the constant matrix. For this purpose, we first let each column of $M$, $M[:,i] \equiv \vec v_i$ ($i \in [1,\ldots,d_{out}]$) be a vertex $v_i$ in a graph. Then, we add the root vertex $v_0$ associated with the zero vector $\vec 0$ to the graph. The distance between two vertices $v_i$ and $v_j$ is defined as the sum of the number of non-zero digits in the vector $\vec v_i+\vec v_j$ or $\vec v_i-\vec v_j$, whichever is lower.

Starting from $v_0$ associated with $\vec 0$, we use Prim's algorithm~\cite{prim1957shortest} to find an approximate Minimum Spanning Tree (MST) of the graph, subject to a maximum depth smaller than or equal to $2^{dc}$ from the root vertex. Each edge in the approximate MST is then translated back into a column vector in $M_1$ of shape $[d_{in},d_{out}]$, with its contribution to the original outputs (-1, 0, or 1) recorded in the second submatrix $M_2$ of shape $[d_{out},d_{out}]$. The contribution can be determined by tracing from the root vertex to the corresponding vertex, where each edge traversed has a non-zero contribution. $M_2$ constructed this way is usually significantly more sparse compared to $M_1$. Both $M_1$ and $M_2$ are then passed to the next stage.

This stage is useful for matrices with correlated columns and a high digit count. For matrices where the columns are rarely correlated, the algorithm would usually produce trivial decomposition, where $M_1$ is a shuffled form of $M$ and $M_2$ is a shuffled identity matrix. As a realistic example would be too large for display, we show a minimal example of decomposing a matrix $M$ of size $3\times 3$:
\begin{equation}
  M = \begin{pmatrix}
    |        & |        & |        \\
    \vec v_1 & \vec v_2 & \vec v_3 \\
    |        & |        & |
  \end{pmatrix}
  = \begin{pmatrix}
    0 & 1 & 3 \\
    1 & 2 & 4 \\
    2 & 3 & 5
  \end{pmatrix}
\end{equation}

\begin{figure}[htbp]
  \centering
  \includegraphics[width=0.7\textwidth]{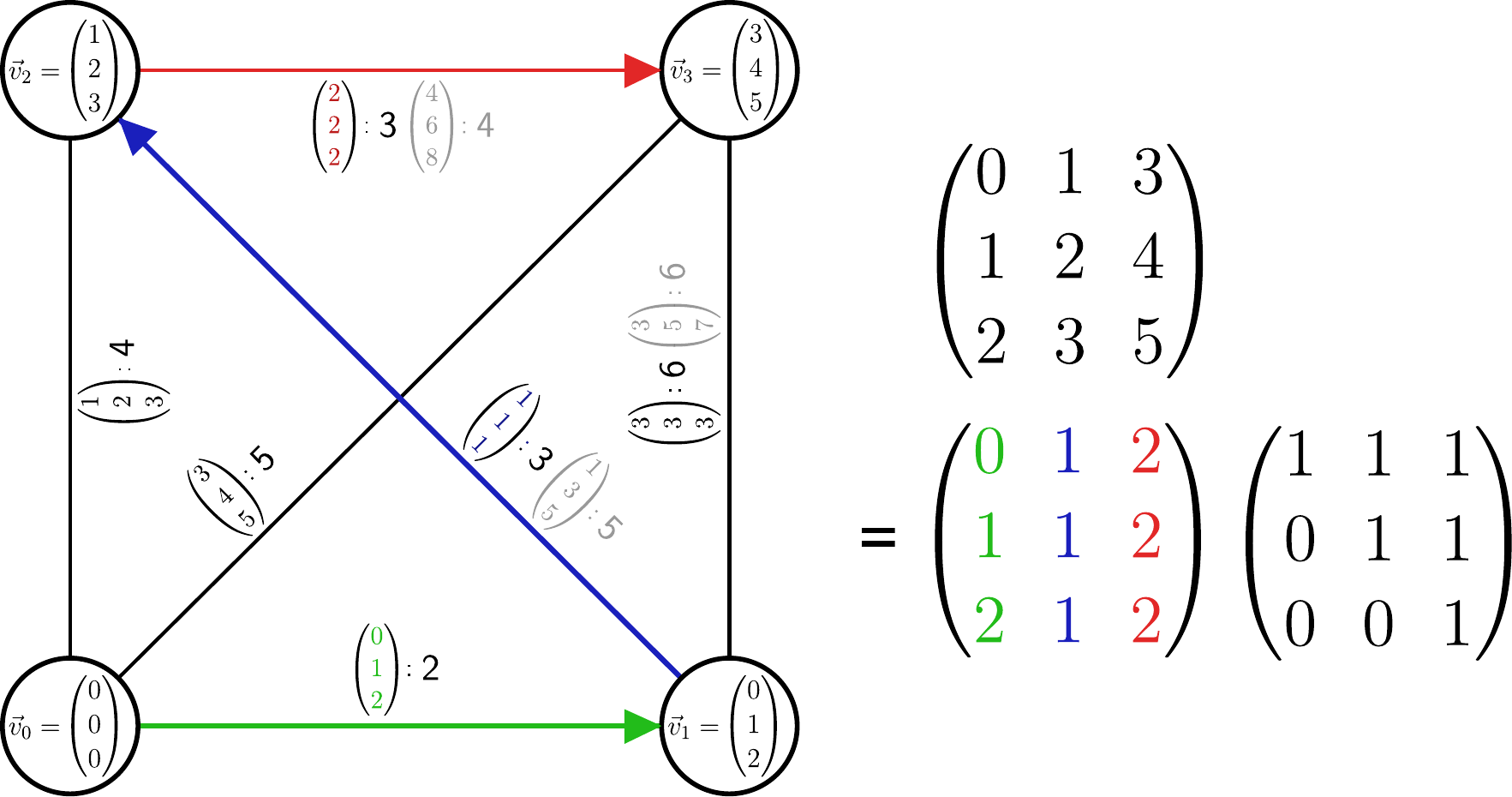}
  \caption{An example of the graph constructed from the constant matrix $M$ in the first stage of the \texttt{da4ml} algorithm without a delay constraint. The graph is constructed with Prim's algorithm, and the MST is shown in colored edges. The root vertex $\vec v_0$ is on the bottom left. The column vectors on the edges are the corresponding vectors that need to be added/subtracted to transform between the two connected vertices. The number after the column vectors is the number of non-zero digits needed to represent the vector, and the smaller of the two is used as the edge weight. The column vectors corresponding to the edges in the MST are stored in $M_1$, and their contributions to $\vec v_1$, $\vec v_2$, and $\vec v_3$ are recorded in $M_2$.
  }
  \label{fig:decompose-example}
  \Description{A diagram connecting a graph structure to its matrix equivalent. On the left, a four-node graph shows a Minimum Spanning Tree highlighted with colored edges, where each edge has a vector and a cost. On the right, an equation demonstrates how an initial 3-by-3 matrix is factored into the product of two matrices derived directly from the structure of the graph on the left.}
\end{figure}

The graph constructed from this matrix is shown in Figure~\ref{fig:decompose-example}. In this particular example, the MST constructed is a chain of $\vec v_0\rightarrow \vec v_1 \rightarrow \vec v_2 \rightarrow \vec v_3$, shown by the colored edges. The corresponding column vectors to these edges are recorded in $M_1$, and their contributions to $\vec v_1$, $\vec v_2$, and $\vec v_3$ are recorded in $M_2$, where $\vec v_1$ is obtained by adding the first edge, $\vec v_2$ is obtained by adding the first two edges, and $\vec v_3$ is obtained by adding all three edges.

\subsection{Second stage: Common Subexpression Elimination}

In the second stage, CSE is applied independently on $M_1$ and $M_2$, and the solutions are concatenated to form the final solution. Common subexpression elimination, CSE, is a compiler optimization that finds identical computations that would otherwise be repeated, computes them once, and reuses the result. For simplicity, we denote the input matrix for the second stage as just $M$. In this stage, $M$ is again normalized into $\hat M$, which has no row or column containing all even numbers except for zeros. The algorithm then converts the input matrix to the CSD representation $M_{expr}\in(-1,0,1)^{d_{in},d_{out},B}$, where $B$ is the span of the powers of the CSD digits (i.e., the difference between the minimal and maximal bit-shifts associated with the CSD digits plus one).

The CSE algorithm starts with the $M_{expr}$ matrix, and a list of implemented values $L_{impl}$ initialized with the elements of the input vector: $L_{impl}=[v_1,v_2,\ldots,v_{d_{in}}]$. These two objects define the state of the algorithm, and the algorithm updates the state iteratively until exhausting all common subexpressions.

For each update step, the algorithm selects a two-term subexpression and implements it. A two-term subexpression is defined as an operation with the general form $a\pm b<< s$, characterized as a four-tuple: the two inputs $a$ and $b$, the sign, and the bit-shift $s$ for the second operand\footnote{The order of $a$ and $b$ is fixed by their location in the matrix in practice for the uniqueness of the four-tuple.}.

Implementing a two-term subexpression consists of two steps. The algorithm first appends the result of the two-term subexpression to the list of implemented values $L_{impl}$. Then, it appends an empty row to $M_{expr}$. For every occurrence of the subexpression in $M_{expr}$, the algorithm sets the corresponding two digits to zero and sets -1 or 1 on the new row at the corresponding power. At all times, the value of $\sum_j v_j \hat M_{ji}$ can be recovered with $\sum_{j,s} 2^{s} \cdot L_{impl}[j]\cdot \hat M_{expr}[j,i,s]$.

In the implementation, we keep a hash table that caches the frequency of all two-term subexpressions to speed up the optimization. While \cite{hcmvm,mod-cmvm} show that selecting the two-term subexpression that minimally conflicts with the other subexpressions could result in lower final adder usage, this approach requires a one-step look-ahead operation when selecting the two-term common subexpression in each update step, which involves a complexity of $\mathcal{O}(|L_{impl}|^2)$. Due to practical considerations for larger matrices, we instead select the most common subexpression to implement for each update step, which can be performed with a complexity of $\mathcal{O}(|L_{impl}|)$. As $|L_{impl}|$ is of the same order of magnitude as the total number of non-zero digits in $M_{expr}$, the additional time complexity for performing the look-ahead operation would be substantial. As suggested in~\cite{hcmvm}, because this only improves the final resource utilization by less than 2\% as measured in adder count, we decided that the additional complexity is not worth the runtime overhead.

In contrast to~\cite{most-common-cmvm}, we also take into account the operands' quantized intervals (i.e., the bitwidths and shift) when choosing the subexpression for each update step. As implied by the cost function in \eqref{eq:cost_adder}, it is preferred to have the two operands with similar bitwidths and shifts. However, directly weighting the frequency of the two-term subexpressions by the total cost is not ideal: it would also count the half-adders used, which are "overheads" as they may unnecessarily increase the accumulator width downstream. Instead, we weight the frequency by the number of overlapping bits between the two operands. This weighting reduces to a constant factor when the input bitwidths are uniform and significantly larger than the constant matrix's bitwidths.

\begin{figure}[htbp]
  \centering
  \includegraphics[width=0.85\textwidth]{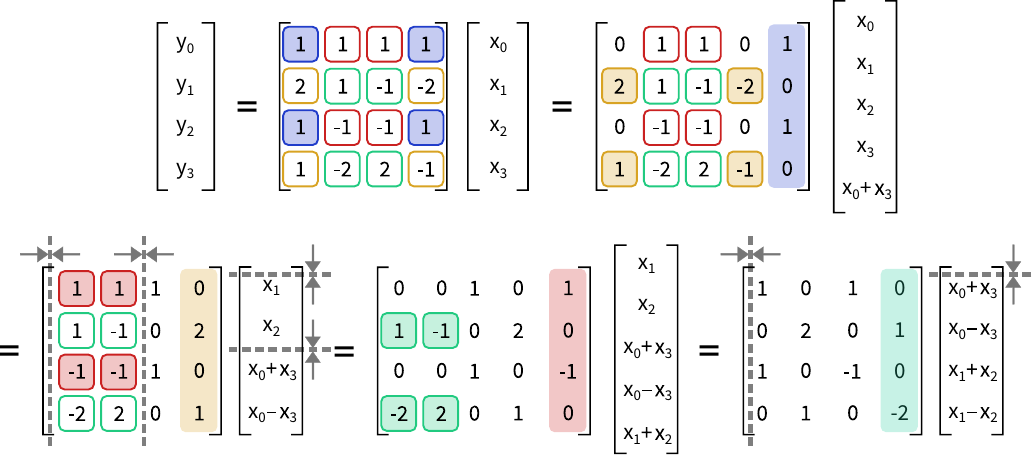}
  \caption{An example of the second stage of the \texttt{da4ml} algorithm. The algorithm identifies three two-term subexpressions, shown in the bounding boxes with the same color. For representation purposes, the matrix shown is a transposed matrix (i.e., $\vec y=M\vec x$ instead of $\vec y^\textrm{T}=\vec x^\mathrm{T}M$). The first subexpression $x_0 + x_3$ has the highest frequency and is implemented first, followed by the other three subexpressions. Frequency weighting by operand bitwidths is not applied in this example for simplicity. In each step, the subexpressions with a colored shade are eliminated. In the last step, the all-zero columns and corresponding elements in the implemented values are omitted.}
  \label{fig:h264-steps}
  \Description{A visualization of the iterative matrix simplification process. A sequence of three matrix-vector equations shows how the algorithm operates. In each step, common subexpressions in the matrix, highlighted by colored shading, are identified and replaced. This progressively transforms the initial dense matrix into a much sparser matrix, visually demonstrating the reduction in operations.}
\end{figure}

\begin{figure}[htbp]
  \centering
  \includegraphics[width=0.9\textwidth]{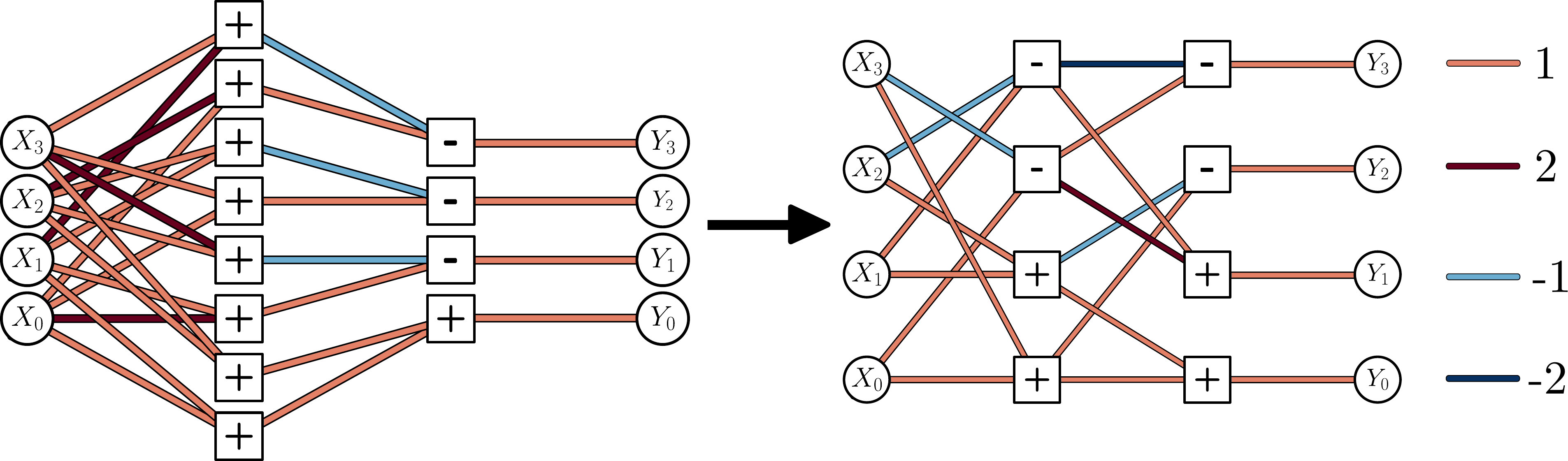}
  \caption{An example of the adder graphs implementing the H.264 constant matrix before and after the optimization. The original adder graph on the left requires 12 adders, while the optimized adder graph on the right only requires 8 adders. Each square node represents an adder/subtractor, and the edges represent the inputs to the adder/subtractor. The color of the edges indicates the sign and the power-pf-two coefficients of the inputs. The circle nodes represent the inputs to the adder graph.}
  \label{fig:h264-before-after.pdf}
  \Description{A before-and-after comparison of two circuit diagrams. The 'before' diagram on the left shows a network of twelve square adder nodes connecting inputs to outputs. The 'after' diagram on the right shows a simplified network for the same function containing only eight adder nodes. This visually illustrates the hardware resource savings achieved by the optimization.}
\end{figure}

We show a minimal example of the second stage in Figure~\ref{fig:h264-steps} with a constant matrix from the H.264 integer transform~\cite{h264}. For presentation purposes, the matrix shown is a transposed matrix (i.e., $\vec y=M\vec x$ instead of $\vec y^\textrm{T}=\vec x^\mathrm{T}M$). In this minimal example, we do not weight the frequency of the subexpressions by the operand bitwidths for simplicity. Also, please note that although the example matrix contains only power-of-two numbers, the algorithm accepts matrices with arbitrary fixed-point elements. The algorithm first identifies three two-term subexpressions, shown in the bounding boxes with the same color. In the first step, the expression $x_0 + x_3$ is implemented, and all of its occurrences are replaced. The algorithm then proceeds to implement the other three subexpressions, $x_0 - x_3$, $x_1 + x_2$, and $x_1 - x_2$. The adder graphs before and after optimization are shown in Figure~\ref{fig:h264-before-after.pdf}, where the original adder graph requiring 12 adders on the left is reduced to 8 adders on the right.

\subsection{Complexity Analysis and Performance}

For complexity analysis, we denote the bitwidth of the constant matrix $M$ as $bw_M$, and the number of inputs and outputs as $d_{in}$ and $d_{out}$, respectively. Let $N = d_{in}\cdot d_{out}\cdot bw_M$.

The complexity of the algorithm is dominated by the second stage. As the first stage is a constrained variation of the Prim's algorithm, the complexity of the first stage is bounded $\mathcal{O}(d_{out}^2)$.

For the second stage, we found that the complexity is dominated by optimizing $M_1$, for which it is similar to optimizing $M$ directly.

Initializing the hash table requires a complexity of $\mathcal{O}\left(d_{out}\cdot(d_{in}\cdot bw_M)^2\right)$ for double iteration over the input dimensions and single iteration over the output expressions. $\mathcal{O}(|L_{impl}|)\sim\mathcal{O}(N)$ update steps are required to reach the final solution. For each step, the complexity is $\mathcal{O}(N)$ to find the most common two-term subexpression over a cached dictionary. Each frequency update also takes $\mathcal{O}(N)$ time, as only single iteration over the input dimensions is required for differential updates.

Hence, the overall asymptotic complexity of the algorithm is expected to be $\mathcal{O}(N^2)$, dominated by the iterative update steps.

In practice, we found the execution time of the algorithm to be asymptotically close to $\mathcal{O}(N^2\cdot \log(N)^2)$ up to $N\sim 10^5$ ($128\times128\times 8$-bit). We postulate that the logarithmic factor is due to the overhead of the hash table or memory allocation/deallocation when appending rows to the $M_{expr}$ matrix.

\section{Implementation and Integration}
\label{sec:impl}

\texttt{da4ml} is implemented as a high-performance Python library and is open-sourced. At its core, \texttt{da4ml} leverages Numba~\cite{numba}, a just-in-time compiler for Python that compiles code to MLIR and then to a binary with LLVM~\cite{llvm} for performance considerations.

\texttt{da4ml} provides two main functionalities: (1) it can be used as a plug-in optimizer for CMVM operations in the popular \texttt{hls4ml} library~\cite{hls4ml}, and (2) it can also be used standalone to generate synthesizable RTL code for most fully parallel neural networks.

\subsection{Integration with \texttt{hls4ml}}

\begin{figure}[htbp]
  \centering
  \includegraphics[width=0.75\textwidth]{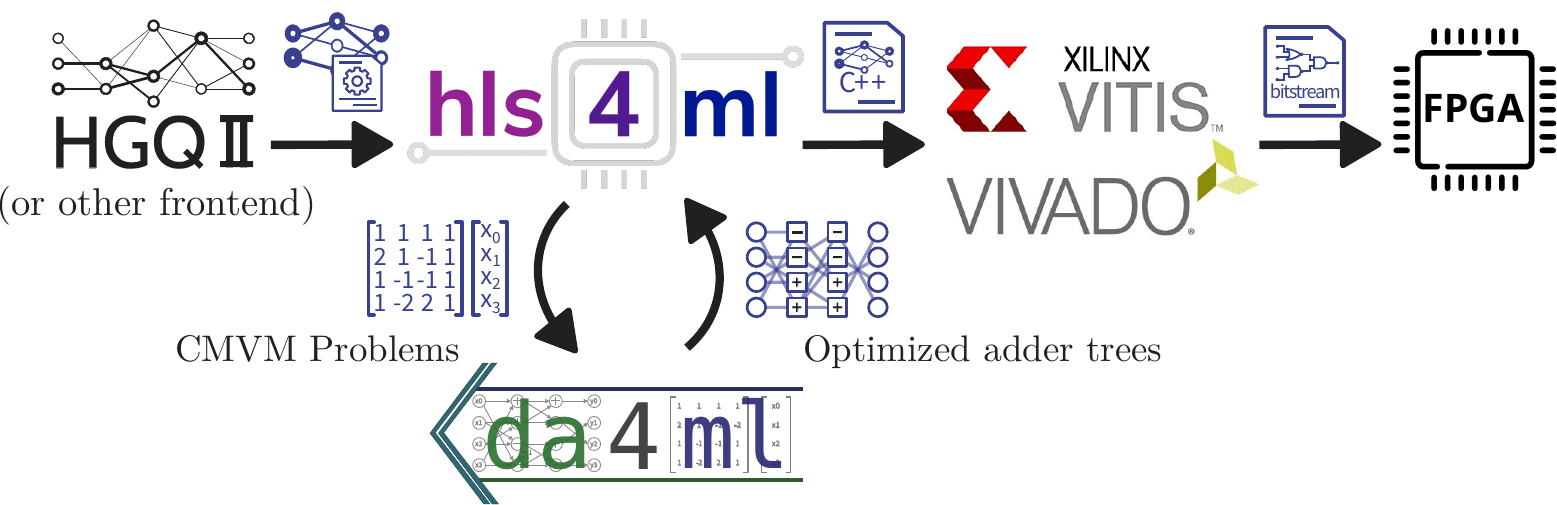}
  \caption{The workflow of using the \texttt{da4ml} library with the \texttt{hls4ml} library. First, the high-level model description (HGQ or another frontend) is converted into HLS C++ by the \texttt{hls4ml} library. The CMVMs involved are then passed to the \texttt{da4ml} library, which generates optimized adder trees tailored for the CMVM operations. The optimized adder trees are then reinserted into the hls4ml-generated code as a drop-in replacement for the default CMVM implementation. Finally, the modified HLS code is compiled by Xilinx Vitis/Vivado to produce the FPGA bitstream. In this workflow, da4ml acts as an intermediate optimization step between \texttt{hls4ml} code generation and FPGA synthesis, with its output directly reducing the latency of the final hardware design.
  }
  \label{fig:workflow}
  \Description{A toolchain integration flowchart using logos of the tools involved. Arrows show the flow starting from a frontend like HGQ into hls4ml. A matrix problem is passed from \texttt{hls4ml} to the \texttt{da4ml} library, which returns an optimized adder tree. The flow continues from \texttt{hls4ml} through Xilinx Vivado to generate a bitstream for an FPGA, illustrating da4ml's role as a plug-in optimizer.}
\end{figure}

Designed for seamless deployment, the library is tightly integrated with the widely adopted \texttt{hls4ml} framework~\cite{hls4ml}. Users can easily enable \texttt{da4ml} by setting the strategy to \texttt{distributed\_arithmetic} for any layer requiring CMVM operations, as shown in Listing~\ref{code:da}. No further user configuration or intervention is required. The overall workflow is shown in Figure~\ref{fig:workflow}. Here, HGQ~\cite{hgq} is a quantization-aware training framework with differentiable quantization, where each weight has its own bitwidth. The trained model is bit-wisely highly sparse, making it a suitable candidate for optimization with \texttt{da4ml}. In this workflow, \texttt{da4ml} generates an optimized adder tree for the CMVM operation in HLS C++ code, which is then used as a drop-in replacement for the default CMVM implementation within the \texttt{hls4ml}.
This drop-in integration enables significant improvements in resource efficiency and latency without sacrificing usability. Currently, the \texttt{Dense}, \texttt{EinsumDense}, and \texttt{Conv1D/Conv2D} layers are supported for the Vivado/Vitis backends, with plans for broader support underway. By bridging algorithmic innovation with production-ready toolchains, da4ml provides a scalable and practical solution for deploying high-throughput, low-latency neural networks on FPGAs.

\begin{lstlisting}[float=htbp,emph={with},emphstyle={\bfseries}, language=Python, caption=Enabling the DA strategy on all supported layers in \texttt{hls4ml} using our framework., label=code:da]
# configuration
hls_config = {'Model': {
    'Precision': <INSERT_MODEL_PRECISION>,
    'ReuseFactor': 1, 
    'Strategy': 'distributed_arithmetic'
}}

# convert, may also be from other frameworks
model_hls = convert_from_keras_model(model, hls_config=hls_config, ...)

model_hls.compile() # Code emission and functional simulator binding
test_output = model_hls.predict(test_data) # Running the simulation

\end{lstlisting}

\subsection{Standalone Code Generation}
\begin{figure}[htbp]
  \centering
  \includegraphics[width=0.75\textwidth]{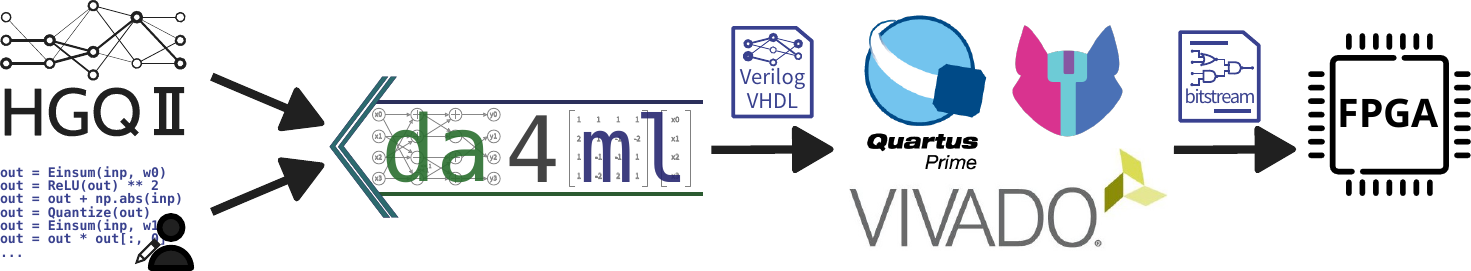}
  \caption{The workflow of the standalone code generation with \texttt{da4ml}. To define a neural network in this workflow, the user needs to apply the desired operations using overloaded numpy operators on symbolic tensors provided by the library. For neural networks defined in HGQ2~\cite{hgq2}, the symbolic tracing process is automated. The library then automatically generates synthesizable RTL code either in fully pipelined or combinational form in Verilog or VHDL.
  }
  \label{fig:da4ml_workflow}
  \Description{A flowchart for the standalone RTL generation workflow. To define a neural network, the user needs to apply the desired operations using overloaded numpy operators on symbolic tensors provided by the da4ml library. For neural networks defined in HGQ2, the symbolic tracing process is automated. The library then automatically generates synthesizable RTL code either in fully pipelined or combinational form in Verilog or VHDL. Verification of the generated RTL designs is provided by Verilator and GHDL integration.}
\end{figure}

\texttt{da4ml} can also work standalone to generate RTL (Verilog or VHDL) designs for most fully parallel neural networks as an alternative to \texttt{hls4ml}, allowing for easier integration into existing RTL workflows. Verilator~\cite{verilator} and GHDL~\cite{ghdl} are integrated for bit-and-cycle-accurate verification of the generated RTL code. Currently, the generated designs must be either fully pipelined with an II of 1 or combinational. At its current stage, \texttt{da4ml} does not support generalized table lookups (e.g., generic activation functions) or stateful operations (e.g., RNNs or LSTMs~\cite{que2021accelerating}), while most common neural network layers (e.g., fully connected, convolutional, pooling, batch normalization) are supported. Users can define neural networks with flexible structures using symbolic tracing by applying the high-level operations with overloaded \texttt{numpy} operators on the symbolic tensors provided by the library, and the library will record the sequence of operations and generate the corresponding RTL code. For models defined in HGQ2~\cite{hgq2}, the symbolic tracing process is automatically performed, as demonstrated in the designs of JEDI-linear~\cite{jedi-linear}.
The standalone RTL generation with \texttt{da4ml} is shown in Figure~\ref{fig:da4ml_workflow}.

When this workflow is used, after the symbolic tracing, the framework lowers the high-level operations into its internal representation, Distributed Arithmetic Instruction Set (DAIS). As its name suggests, DAIS is a low-level, instruction-set-like representation that consists of only a few operations, and the program will be produced in a static single assignment form. Each DAIS program directly describes a combinational circuit, and emitting RTL code from DAIS can be achieved by simply mapping each DAIS operation to its corresponding RTL module.

Pipelining is achieved by inserting registers between each DAIS operation based on the estimated delay of each operation. When the estimated delay exceeds the user defined threshold, registers are inserted to break the long combinational path. By default, each adder is assumed to have a delay of 1 unit regardless of the input bitwidths, while the exact mapping is user-configurable. The pipelining algorithm is greedy and local, and it does not perform any global optimization. However, as the majority of the delay in FPGA designs is due to routing, this simple approach is usually sufficient to meet timing with a reasonable margin at the cost of higher register usage than the HLS counterpart. Since the pipelining is not timing driven, it is recommended to enable the backend synthesis tool's retiming functionality for best performance.

\begin{lstlisting}[float=htbp,emph={with},emphstyle={\bfseries}, language=Python, caption=Defining a model directly using \texttt{da4ml}., label=code:da_rtl]
inp, out = trace_model(hgq2_model) # Automatic symbolic tracing
comb_logic = comb_trace(inp, out) # Lowering to DAIS

# Defining the RTL model and piplining
rtl_model = RTLModel(comb_logic, 'prj_name', '/output/path', flavor='verilog', latency_cutoff=5)

rtl_model.compile() # Code emission and Verilator binding
test_output = rtl_model.predict(test_data) # Running the simulation
\end{lstlisting}

\section{Experiments}
\label{sec:experiments}

This section first evaluates the proposed \texttt{da4ml} algorithm on random matrices and compares the high-level metrics with the state-of-the-art \hcmvm algorithm. We then evaluate the performance of the proposed algorithm on realistic neural networks consisting CMVM operations. We perform both the HLS version and hardware-description-language (HDL) version of the algorithm. The HLS version is implemented with the \texttt{hls4ml} library~\cite{hls4ml}, and the HDL version is implemented standalone with the \texttt{da4ml} library.

\newcommand{\mcol}[3]{\multicolumn{#1}{#2}{#3}}
\subsection{Random Matrices}

We evaluate the performance of the proposed algorithm on random $m\times m$ matrices. In this subsection, to facilitate comparison with the \hcmvm algorithm, we adopt a convention from~\cite{hcmvm} where a $bw$-bit random matrix is generated by sampling integers uniformly from $[2^{bw-1}+1, 2^{bw}-1]$.

\label{sec:random_mat}
\begin{table}[htbp]
  \begin{adjustbox}{width=1.0\textwidth,center=\textwidth}
    \begin{tabular}{l|ccc|ccc|ccc|ccc|ccc|cc}
      \toprule
                          & \mcol{6}{c}{$dc=-1$ (no delay constraint)} & \mcol{6}{|c}{$dc=0$} & \mcol{5}{|c}{$dc=2$}                                                                                                                                                        \\
      \hline
                          &                                            & da4ml                &                      &       & \hcmvm~\cite{hcmvm}              &          &       & da4ml &          &       &
      \hcmvm~\cite{hcmvm} &                                            &                      & da4ml                &       & \mcol{2}{c}{\hcmvm~\cite{hcmvm}}                                                                                                             \\
      \hline

      N                   & depth                                      & adder                & cpu [ms]             & depth & adder                            & cpu [ms] & depth & adder & cpu [ms] & depth & adder & cpu [ms] & depth & adder & cpu [ms] & depth & adder \\
      \midrule
      2                   & 3.3                                        & 8.7                  & 0.1                  & 4.4   & 8.2                              & 1.0e1    & 3.1   & 9.9   & 0.1      & 3.1   & 8.8   & 1.0e1    & 3.3   & 8.7   & 0.1      & 3.7   & 8.2   \\
      4                   & 6.1                                        & 29.3                 & 0.3                  & 7.8   & 27.6                             & 4.8e2    & 4.1   & 37.   & 0.3      & 4.1   & 32.1  & 4.7e2    & 5.9   & 30.   & 0.3      & 5.7   & 28.1  \\
      6                   & 8.4                                        & 59.                  & 0.6                  & 10.   & 57.3                             & 3.3e3    & 5.    & 77.8  & 0.8      & 5.    & 66.8  & 3.8e3    & 6.7   & 62.6  & 0.6      & 7.    & 58.2  \\
      8                   & 9.4                                        & 98.                  & 1.3                  & 11.9  & 96.3                             & 1.5e4    & 5.1   & 130.9 & 2.       & 5.1   & 117.2 & 1.7e4    & 7.    & 102.3 & 1.4      & 7.1   & 99.5  \\
      1                   & 10.8                                       & 146.6                & 2.7                  & 13.2  & 143.5                            & 5.4e4    & 6.    & 195.6 & 4.2      & 6.    & 157.7 & 8.2e4    & 7.8   & 152.8 & 2.8      & 8.    & 146.9 \\
      12                  & 11.6                                       & 203.6                & 4.8                  & 14.6  & 200.4                            & 1.7e5    & 6.    & 271.8 & 7.9      & 6.    & 241.6 & 1.7e5    & 8.    & 214.9 & 5.2      & 8.    & 206.8 \\
      14                  & 12.3                                       & 269.3                & 8.3                  & 15.5  & 264.3                            & 4.8e5    & 6.    & 358.5 & 14.1     & 6.    & 324.  & 4.2e5    & 8.    & 279.2 & 8.9      & 8.    & 274.8 \\
      16                  & 13.                                        & 343.4                & 13.3                 & 16.3  & 338.3                            & 1.2e6    & 6.    & 456.  & 22.5     & 6.    & 423.2 & 9.9e5    & 8.    & 358.7 & 14.9     & 8.    & 353.3 \\
      \bottomrule
    \end{tabular}
  \end{adjustbox}
  \caption{Comparison of the \texttt{da4ml} algorithm with the \hcmvm~algorithm on random matrices. ``depth'' is the adder depth (longest path counted in number of adders from input to output), ``adder'' is the total number of adders or subtractor used, and ``cpu [ms]'' is the wall-clock time in milliseconds on CPU with a single thread. The CPU time is measured on a single thread of an Intel i7-13700 CPU at 5.10 GHz for \texttt{da4ml}, and on a single thread of an Intel Xeon at 2.33 GHz for \hcmvm, as reported by Ref.~\cite{hcmvm}. ``dc'' is the delay constraint.}
  \label{tab:random_mat}
\end{table}

Table~\ref{tab:random_mat} shows the comparison of the \texttt{da4ml} algorithm with the \hcmvm~algorithm on random matrices with 8-bits under different delay constraints.
For each configuration, we report the adder depth, total number of adders, and runtime on CPU. Results are shown for three delay constraint ($dc$, see Table~\ref{tab:parameters}) settings: no constraint, strict delay constraint ($dc = 0$), and moderate delay constraint ($dc = 2$). The CPU time is measured on a single thread of an Intel i7-13700 CPU at 5.10 GHz for \texttt{da4ml}, and on a single thread of an Intel Xeon at 2.33 GHz for \hcmvm, as reported by the authors.
The table shows the number of adders used in the optimized design of both algorithms under the same delay constraints, and the adder depth of both. When delay constraint is not exactly 0 (strictly minimal adder depth is required), the proposed algorithm has $\sim 2\%$ overhead in adder counts compared to the \hcmvm~algorithm. When strict delay constraint is required, the proposed algorithm has a larger overhead of $\sim 8\%$ in adder counts. Please note that this slight trade-off in resource usage yields dramatic improvements in compilation speed.

With respect to runtime, the \texttt{da4ml} algorithm is significantly more efficient, being $\sim 10^5$ times faster than the \hcmvm~algorithm with a moderate matrix size of $16\times 16$. The difference would be more significant for larger matrices given the difference in asymptotic complexities. Although there is a generation gap between the two CPU processors, the magnitude of the observed speedup far exceeds what can be attributed solely to CPU differences, highlighting the efficiency and scalability of our framework.

\begin{figure}[htbp]
  \centering
  \includegraphics[width=0.45\textwidth]{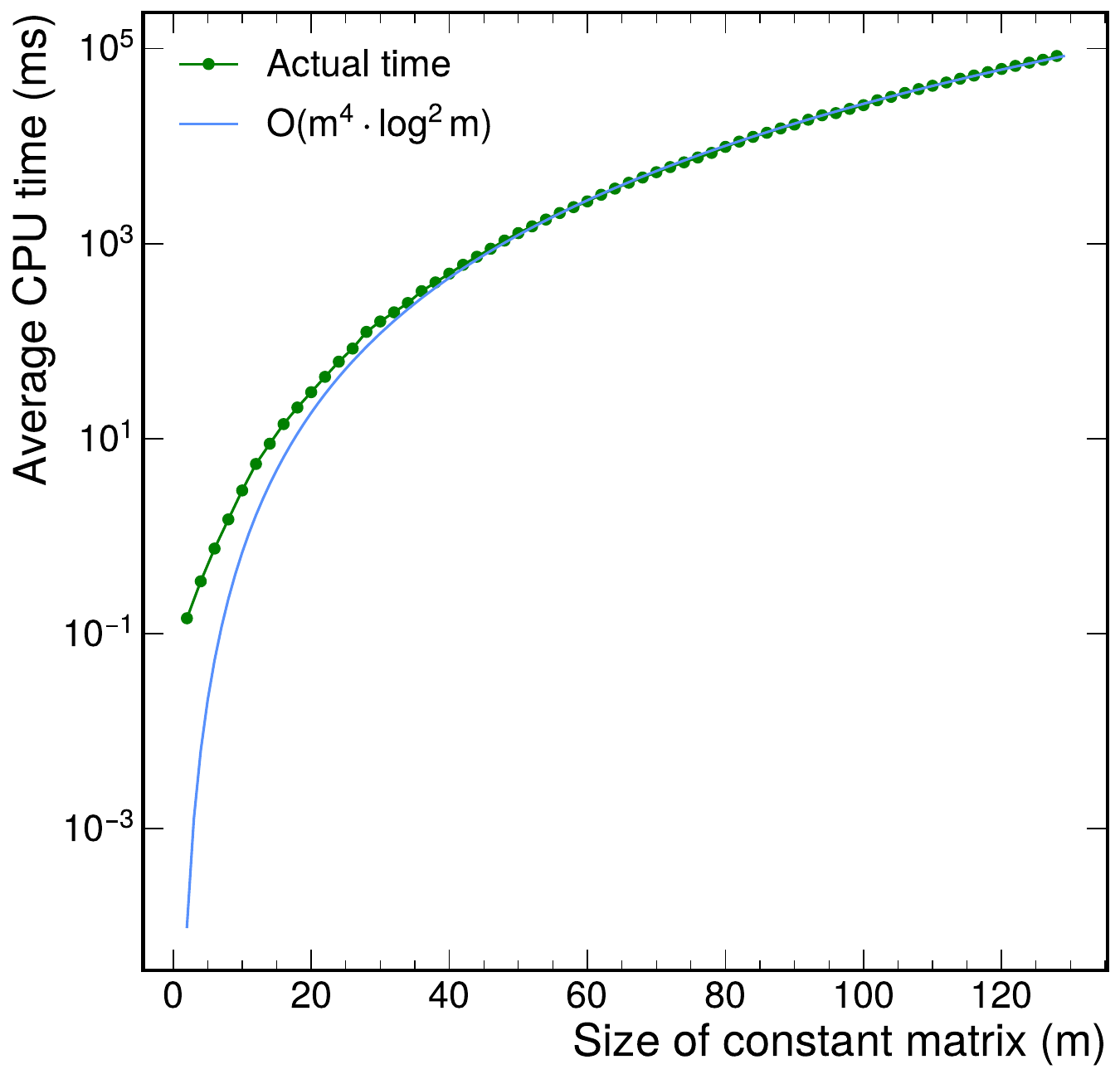}
  \caption{Computation time of the \texttt{da4ml} algorithm on random matrices with different sizes. The asymptotic complexity is $\mathcal{O}(N^2\cdot \log(N)^2)$, where $N\sim d_{in}\cdot d_{out}\cdot bw_M = m^2 \cdot bw_M$. The $\log(N)^2$ factor was found empirically.}
  \label{fig:rand_mat_time}
  \Description{A log-linear plot of algorithm computation time versus matrix size. The horizontal axis is the size of the constant matrix from 0 to 120. The vertical axis is the average CPU time in milliseconds on a logarithmic scale. A line showing the measured "Actual time" curves steeply upwards and is closely matched by a fitted complexity curve labeled O of m to the fourth power times log squared of m.}
\end{figure}

We further show extended runtime results for larger matrices in Figure~\ref{fig:rand_mat_time}, including the computation time of the \texttt{da4ml} algorithm on random matrices with up to $128\times 128$ elements with 8 bits, as well as the time expected from the asymptotic complexity of the algorithm. These results demostrate that the \texttt{da4ml} algorithm may handle reasonably complex CMVM problems that may be implemented on FPGAs unrolled in reasonable time.

We show the post-synthesis results of the proposed \texttt{da4ml} algorithm on random matrices with 8 bits and 4 bits, shown in Table~\ref{tab:random_mat_synth_8bit} and Table~\ref{tab:random_mat_synth_4bit}, respectively. The results shown in this section are obtained after HLS synthesis with Vitis HLS 2023.2 and out-of-context synthesis and place-and-route with the Vivado 2023.2 backend. The target FPGA is \texttt{xcvu13p-flga2577-2-e}. To quantify the logic delay, the designs are synthesized with a latency of one clock cycle, where the CMVM logic is a combinational logic block sandwiched between two layers of registers. For each matrix size, we use the same random constant matrix for the three delay constraints: 0, 2, and -1 (no constraint), marked as DC in the tables.

The baseline for comparison is the latency-optimized implementation in \texttt{hls4ml}, where an unrolled double for-loop is used to implement the CMVM operation with a fixed-point multiplication-accumulation operation for HLS. The inputs for all designs are 8-bit signed integers.

As shown in Table~\ref{tab:random_mat_synth_8bit}, the \texttt{da4ml} approach completely avoids using DSP blocks. In cases where the baseline also relies on LUTs for multiplication (e.g., Table~\ref{tab:random_mat_synth_4bit} or the $8\times8$ matrix in Table~\ref{tab:random_mat_synth_8bit}), \texttt{da4ml} can consistently reduce LUT usage by approximately half.
When the baseline utilizes DSPs, the LUT usage is comparable, though it can be significantly reduced by relaxing the delay constraint. For latency, the critical path of the \texttt{da4ml} designs can be either shorter or longer than the baseline, depending on the chosen bitwidth and delay constraint.

\begin{table*}[htbp]
  \begin{tabular}{lcc|ccccc}
    \toprule
    Strategy & DC & Matrix Size   & LUT   & DSP  & FF    & Latency [ns] & adders  \\
    \midrule
    latency  & -  & $8\times 8$   & 3193  & 0    & 645   & 2.21         & (211)   \\
    DA       & 0  & $8\times 8$   & 1570  & 0    & 654   & 1.97         & 123     \\
    DA       & 2  & $8\times 8$   & 1214  & 0    & 420   & 2.62         & 97      \\
    DA       & -1 & $8\times 8$   & 1200  & 0    & 412   & 3.14         & 93      \\
    \hline
    latency  & -  & $16\times 16$ & 4319  & 212  & 2301  & 3.05         & (845)   \\
    DA       & 0  & $16\times 16$ & 5281  & 0    & 1947  & 2.53         & 436     \\
    DA       & 2  & $16\times 16$ & 4545  & 0    & 1618  & 2.66         & 361     \\
    DA       & -1 & $16\times 16$ & 4321  & 0    & 1283  & 4.11         & 349     \\
    \hline
    latency  & -  & $32\times 32$ & 17666 & 807  & 4886  & 4.43         & (3501)  \\
    DA       & 0  & $32\times 32$ & 18807 & 0    & 6408  & 3.14         & 1591    \\
    DA       & 2  & $32\times 32$ & 15427 & 0    & 4494  & 3.52         & 1263    \\
    DA       & -1 & $32\times 32$ & 14866 & 0    & 4643  & 5.71         & 1228    \\
    \hline
    latency  & -  & $64\times 64$ & 70821 & 2897 & 18969 & 5.63         & (14089) \\
    DA       & 0  & $64\times 64$ & 66864 & 0    & 21503 & 4.65         & 5715    \\
    DA       & 2  & $64\times 64$ & 63852 & 0    & 20509 & 4.91         & 5293    \\
    DA       & -1 & $64\times 64$ & 52198 & 0    & 16214 & 6.49         & 4428    \\
    \bottomrule
  \end{tabular}
  \caption{
    Resource utilization and latency of the \texttt{da4ml} algorithm on random matrices with different sizes and delay constraints. Strategy ``latency'' is the latency-optimized algorithm in \texttt{hls4ml}, and ``DA'' is the optimized implementation with the \texttt{da4ml} algorithm. The delay constraint is set to 0, 2, or -1, where -1 means no delay constraint. The matrix size is $d_{in}\times d_{out}$, and the bitwidth is 8. The input bitwidth is 8 in all cases.
  }
  \label{tab:random_mat_synth_8bit}
\end{table*}

\begin{table*}[htbp]
  \begin{tabular}{lcc|ccccc}
    \toprule
    Strategy & DC & Matrix Size   & LUT   & DSP & FF    & Latency [ns] & adders \\
    \midrule
    latency  & -  & $8\times 8$   & 1241  & 0   & 459   & 1.77         & (124)  \\
    DA       & 0  & $8\times 8$   & 885   & 0   & 375   & 1.70         & 71     \\
    DA       & 2  & $8\times 8$   & 721   & 0   & 278   & 1.86         & 55     \\
    DA       & -1 & $8\times 8$   & 644   & 0   & 275   & 2.48         & 52     \\
    \hline
    latency  & -  & $16\times 16$ & 4538  & 0   & 1585  & 1.88         & (529)  \\
    DA       & 0  & $16\times 16$ & 2922  & 0   & 1084  & 2.30         & 269    \\
    DA       & 2  & $16\times 16$ & 2268  & 0   & 745   & 2.34         & 195    \\
    DA       & -1 & $16\times 16$ & 2126  & 0   & 699   & 3.27         & 178    \\
    \hline
    latency  & -  & $32\times 32$ & 13550 & 0   & 3761  & 2.44         & (2108) \\
    DA       & 0  & $32\times 32$ & 10658 & 0   & 3721  & 2.78         & 927    \\
    DA       & 2  & $32\times 32$ & 7452  & 0   & 2465  & 2.93         & 653    \\
    DA       & -1 & $32\times 32$ & 7339  & 0   & 2442  & 3.62         & 625    \\
    \hline
    latency  & -  & $64\times 64$ & 47274 & 0   & 14652 & 3.10         & (8724) \\
    DA       & 0  & $64\times 64$ & 38205 & 0   & 11711 & 3.37         & 3408   \\
    DA       & 2  & $64\times 64$ & 26715 & 0   & 6026  & 4.18         & 2371   \\
    DA       & -1 & $64\times 64$ & 25493 & 0   & 8279  & 5.09         & 2255   \\
    \bottomrule
  \end{tabular}
  \caption{
    Resource utilization and latency of the \texttt{da4ml} algorithm on random matrices with different sizes and delay constraints. Strategy ``latency'' is the latency-optimized algorithm in \texttt{hls4ml}, and ``DA'' is the optimized implementation with the \texttt{da4ml} algorithm. The delay constraint is set to 0, 2, or -1, where -1 means no delay constraint. The matrix size is $d_{in}\times d_{out}$, and the bitwidth is 4. The input bitwidth is 8 in all cases.
  }
  \label{tab:random_mat_synth_4bit}
\end{table*}

\subsection{Realistic Neural Networks}

We show the performance of the \texttt{da4ml} algorithm on realistic neural networks with CMVM operations. The networks are trained with the \texttt{HGQ}~\cite{hgq} library. It is worth noting these neural networks trained with \texttt{HGQ} have heterogeneous bitwidths within one CMVM operation, i.e., there is a high bit-wise sparsity in the constant matrix and the inputs. Due to the extremely heterogeneous bitwidths, \texttt{da4ml}'s benefit is less significant than the random matrix case. For each table, the same neural network is used and different precisions are obtained with different quantization levels. Based on the results from the previous section, we found that a delay constraint of 2 is a good trade-off between resource utilization and latency, and we use it for all evaluations in this section.

For all experiments in this section, the results shown are obtained from out-of-context synthesis and place-and-route with Vivado 2023.2~\cite{vivado}. When \texttt{hls4ml} is used and HLS synthesis is required, we use Vitis HLS 2023.2~\cite{vitis} to generate the RTL designs. In this case, we do not specify the target latency and let Vitis determine the required pipeline stages to meet timing. In all cases, all FPGA resources and $\mathrm{F}_\mathrm{max}$ are reported from the Vivado post routing utilization and timing reports. We target the AMD UltraScale+ FPGA VU13P (\texttt{xcvu13p-flga2577-2-e}) for all experiments in this section unless otherwise specified.

\subsubsection{High-level Feature Jet Tagging Network}

We show the out-of-context results after place-and-route for the high-level feature jet tagging network~\cite{hls4ml} on the VU13P FPGA. The network is a fully-connected neural network with 4 dense layers of sizes $16\rightarrow64\rightarrow32\rightarrow16\rightarrow16\rightarrow5$. Table~\ref{tab:jsc_200MHz} and Table~\ref{tab:jsc_1GHz} show the resource utilization and latency of the network synthesized against target clock frequencies of 200 MHz and 1 GHz, respectively. All designs are fully pipelined with an II of 1 clock cycle.

When the target clock frequency is 200 MHz, the designs optimized with the \texttt{da4ml} algorithm always meet the timing constraints, whereas \texttt{hls4ml}'s latency-optimized designs fail to meet timing for the model with 76.9\% accuracy. In all cases, \texttt{da4ml} reduces LUT consumption by $\sim10$\% for this network. To obtain the maximum frequency, we resynthesized the designs with a target clock frequency of 1 GHz. Under this setting, the designs optimized with the \texttt{da4ml} algorithm are pipelined with fewer stages and achieve $\mathrm{F}_\mathrm{max}$ values similar to or higher than the baseline \texttt{hls4ml} designs. As more registers are inserted, LUT fusion and sharing are less effective, and the LUT utilization is higher than for the designs with a 200 MHz target clock frequency. The DSP utilization is reduced to 0 in all cases with \texttt{da4ml}.

\newcommand{\trow}[1]{\multirow{2}{*}{#1}}
\begin{table*}[htbp]
  \begin{tabular}{lc|cccc cc}
    \toprule
    Strategy & Accuracy      & Latency [cycles] & LUT    & DSP & FF    & $\mathrm{F}_\mathrm{max}$ [MHz] & adders \\
    \midrule
    Latency  & \trow{76.9\%} & 4 (21.6 ns)      & 13,258 & 55  & 1,497 & 185.5                           & (1316) \\
    DA       &               & 4 (18.9 ns)      & 12,250 & 0   & 1,502 & 211.5                           & 992    \\
    \hline
    Latency  & \trow{76.6\%} & 3 (14.7 ns)      & 7,502  & 27  & 900   & 203.6                           & (711)  \\
    DA       &               & 3 (13.3 ns)      & 6,869  & 0   & 966   & 226.0                           & 586    \\
    \hline
    Latency  & \trow{76.5\%} & 3 (14.3 ns)      & 6,690  & 30  & 845   & 210.2                           & (657)  \\
    DA       &               & 3 (13.1 ns)      & 6,005  & 0   & 871   & 229.6                           & 545    \\
    \hline
    Latency  & \trow{76.3\%} & 3 (14.0 ns)      & 5,209  & 15  & 799   & 215.0                           & (495)  \\
    DA       &               & 3 (12.8 ns)      & 4,728  & 0   & 786   & 234.1                           & 422    \\
    \hline
    Latency  & \trow{76.0\%} & 3 (13.4 ns)      & 3,498  & 22  & 639   & 223.2                           & (338)  \\
    DA       &               & 3 (12.3 ns)      & 3,308  & 0   & 661   & 244.8                           & 291    \\
    \hline
    Latency  & \trow{75.9\%} & 3 (13.5 ns)      & 3,043  & 15  & 621   & 221.5                           & (297)  \\
    DA       &               & 3 (12.8 ns)      & 2,878  & 0   & 601   & 234.0                           & 256    \\
    \bottomrule
  \end{tabular}
  \caption{Resource utilization and latency of the high-level feature jet tagging networks with and without \texttt{da4ml} generated with \texttt{hls4ml}, marked with ``DA'' and ``Latency'' for strategy, respectively.
    The FPGA part is \texttt{xcvu13p-flga2577-2-e} with 200 MHz target clock frequency. Delay constraint is set to 2 for the \texttt{da4ml} optimized designs for each CMVM operation. II is 1 for all designs shown.
    The accuracy stands for the top-1 classification accuracy on the test set of the dataset, and the difference in accuracy between different models are due to different quantization levels.
  }
  \label{tab:jsc_200MHz}
\end{table*}

\begin{table*}[htbp]
  \begin{tabular}{lc|cccccc}
    \toprule
    Strategy & Accuracy      & Latency [cycles] & LUT    & DSP & FF     & $\mathrm{F}_\mathrm{max}$ [MHz] & adders \\
    \midrule
    Latency  & \trow{76.9\%} & 42 (57.6 ns)     & 16,081 & 57  & 26,484 & 729.4                           & (1316) \\
    DA       &               & 31 (44.1 ns)     & 12,682 & 0   & 19,056 & 702.2                           & 992    \\
    \hline
    Latency  & \trow{76.6\%} & 32 (46.1 ns)     & 9,347  & 28  & 15,216 & 694.0                           & (711)  \\
    DA       &               & 26 (37.4 ns)     & 7,392  & 0   & 11,462 & 695.9                           & 586    \\
    \hline
    Latency  & \trow{76.5\%} & 35 (67.2 ns)     & 8,548  & 30  & 14,418 & 520.8                           & (657)  \\
    DA       &               & 25 (35.4 ns)     & 6,448  & 0   & 10,109 & 707.2                           & 545    \\
    \hline
    Latency  & \trow{76.3\%} & 33 (51.8 ns)     & 6,667  & 15  & 11,184 & 637.3                           & (495)  \\
    DA       &               & 22 (30.2 ns)     & 5,019  & 0   & 7,682  & 729.4                           & 422    \\
    \hline
    Latency  & \trow{76.0\%} & 33 (48.3 ns)     & 4,471  & 22  & 7,279  & 683.5                           & (338)  \\
    DA       &               & 23 (32.6 ns)     & 3,602  & 0   & 5,693  & 706.2                           & 291    \\
    \hline
    Latency  & \trow{75.9\%} & 31 (42.8 ns)     & 4,005  & 17  & 6,305  & 723.6                           & (297)  \\
    DA       &               & 21 (24.9 ns)     & 2,908  & 0   & 4,720  & 844.6                           & 256    \\
    \bottomrule
  \end{tabular}
  \caption{
    Resource utilization and latency of the high-level feature jet tagging network with and without \texttt{da4ml} generated with \texttt{hls4ml}, marked with ``DA'' and ``Latency'' for strategy, respectively.
    The FPGA part is \texttt{xcvu13p-flga2577-2-e} with 1 GHz target clock frequency. Delay constraint is set to 2 for the \texttt{da4ml} optimized designs for each CMVM operation. II is 1 for all designs shown. The accuracy stands for the top-1 classification accuracy on the test set of the dataset, and the difference in accuracy between different models are due to different quantization levels. Note that the hardware metrics shown are reported by Vivado after out-of-context synthesis and place-and-route, which may differ from the actual maximum frequency (Fmax) on physical hardware.
  }
  \label{tab:jsc_1GHz}
\end{table*}

\subsubsection{SVHN Classification Network}
\begin{figure}[htbp]
  \centering
  \includegraphics[width=0.8\textwidth]{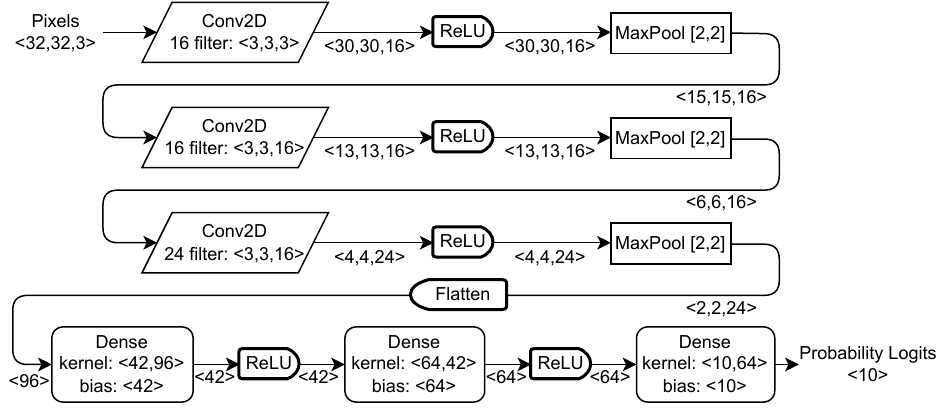}
  \caption{The architecture of the SVHN classification network~\cite{hls4ml}. The network is a LeNet-like~\cite{lenet} network from~\cite{fast_cnn}. The trained networks are taken from~\cite{hgq}.}
  \label{svhn}
  \Description{A block diagram showing data flow through a LeNet-like convolutional neural network. The architecture consists of three sequential blocks, each containing a Conv2D, ReLU, and MaxPool layer, which progressively reduce the data's spatial dimensions. This is followed by a flatten operation and a classification head made of three dense layers. Data dimensions are specified at each stage.}
\end{figure}

We show the out-of-context results after Vivado place \& route for the SVHN classification network~\cite{hls4ml} on the VU9P (\texttt{xcvu9p-flga2104-2L-e}) FPGA. The network is a LeNet-like~\cite{lenet} convolutional network with dense classification head from~\cite{fast_cnn}, and the architecture is shown in Figure~\ref{svhn}. The trained networks are taken from~\cite{hgq}. In contrast to other networks shown in this work, these networks are not fully pipelined, and many CMVM kernels are applied to different regions of the inputs multiple times in one forward pass. The designs are synthesized with Vivado HLS 2020.1 due to inconsistent behaviors of the \texttt{DATAFLOW} pragma used in \texttt{hls4ml}. More details for these networks can be found in~\cite{hgq}. The target clock frequency is set to 200 MHz, and the delay constraint is set to 2 for each CMVM operation. The results are shown in Table~\ref{tab:svhn_200MHz}. From the table, it can be seen that the designs optimized with \texttt{da4ml} can reduce the LUT utilization by $\sim  1/4$, while removing all DSPs. No timing violations are observed in all cases.

\begin{table*}[htbp]
  \begin{tabular}{lc|ccccccc}
    \toprule
    Strategy & Accuracy      & Latency [cycles]   & LUT    & DSP & FF     & $\mathrm{F}_\mathrm{max}$ [MHz] & II [cycles] & adders \\
    \midrule
    Latency  & \trow{93.9\%} & 1,050 (5,058.9 ns) & 69,407 & 58  & 27,853 & 207.6                           & 1,029       & (3022) \\
    DA       &               & 1,045 (5,159.2 ns) & 53,425 & 0   & 20,048 & 202.6                           & 1,029       & 2066   \\
    \hline
    Latency  & \trow{93.1\%} & 1,061 (5,026.0 ns) & 47,314 & 30  & 20,582 & 211.1                           & 1,029       & (2567) \\
    DA       &               & 1,045 (4,952.3 ns) & 36,739 & 0   & 15,491 & 211.0                           & 1,029       & 1783   \\
    \hline
    Latency  & \trow{91.9\%} & 1,058 (5,211.7 ns) & 40,032 & 15  & 18,087 & 203.0                           & 1,029       & (2218) \\
    DA       &               & 1,045 (5,020.2 ns) & 31,267 & 0   & 14,293 & 208.2                           & 1,029       & 1551   \\
    \hline
    Latency  & \trow{90.8\%} & 1,059 (4,868.2 ns) & 34,435 & 13  & 17,261 & 217.5                           & 1,029       & (6271) \\
    DA       &               & 1,045 (5,041.1 ns) & 28,078 & 0   & 13,038 & 207.3                           & 1,029       & 3949   \\
    \hline
    Latency  & \trow{89.9\%} & 1,056 (4,819.6 ns) & 30,766 & 10  & 15,205 & 219.1                           & 1,029       & (4633) \\
    DA       &               & 1,045 (4,993.0 ns) & 24,947 & 0   & 12,261 & 209.3                           & 1,029       & 3004   \\
    \hline
    Latency  & \trow{88.8\%} & 1,056 (5,120.5 ns) & 27,982 & 6   & 14,736 & 206.2                           & 1,029       & (3545) \\
    DA       &               & 1,045 (5,134.1 ns) & 23,609 & 0   & 12,021 & 203.5                           & 1,029       & 2364   \\
    \bottomrule
  \end{tabular}
  \caption{Resource utilization and latency of the SVHN classification network with and without \texttt{da4ml} generated with \texttt{hls4ml}, marked with ``DA'' and ``Latency'' for strategy, respectively.
    The FPGA part is \texttt{xcvu9p-flga2104-2L-e} with 200 MHz target clock frequency. Delay constraint is set to 2 for the \texttt{da4ml} optimized designs for each CMVM operation. The accuracy stands for the top-1 classification accuracy on the test set of the dataset, and the difference in accuracy between different models are due to different quantization levels.
  }
  \label{tab:svhn_200MHz}
\end{table*}

\subsubsection{Muon Tracking Network}
\begin{figure}[htbp]
  \centering
  \includegraphics[width=0.8\textwidth]{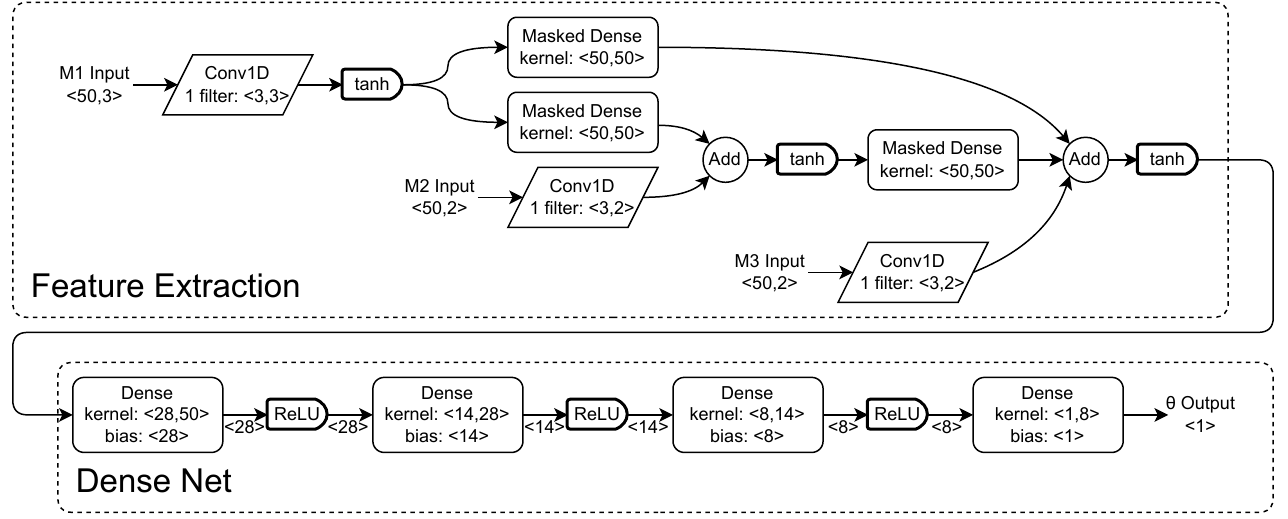}
  \caption{The architecture of the Muon Tracking network~\cite{tgc}. The network is a multi-stage network with mainly dense layers. The masked dense layers refers to dense layers with specific sparsity pattern enforced on the weights, and please refer to~\cite{tgc} for the detailed implementation. The trained networks are taken from~\cite{hgq}.}
  \label{tgc}
  \Description{A diagram of a multi-stage neural network architecture. The top part, labeled "Feature Extraction," shows three parallel branches processing inputs with Conv1D layers, with their outputs being summed together. The bottom part, labeled "Dense Net," takes the extracted features and feeds them through a sequential chain of four dense layers to produce the final single output.}
\end{figure}

The results after place-and-route for the Muon Tracking network~\cite{hls4ml} on the VU13P FPGA are shown in Table~\ref{tab:tgc_160MHz}. The network is a multi-stage network with mainly dense layers, and the architecture is shown in Figure~\ref{tgc}. The trained networks are taken from~\cite{hgq}, and all designs are synthesized with the HLS workflow. The target clock frequency is set to 160 MHz to match the previous work~\cite{tgc}, and the delay constraint is set to 2 for each CMVM operation optimized. As the inputs of this network are all 1-bit, we do not apply \texttt{da4ml} to the initial convolutional layers, as they can be efficiently implemented using conditional accumulation logic. All implementations shown have an II of 1 clock cycle.

Compared to the latency-optimized baseline from \texttt{hls4ml}, the \texttt{da4ml}-optimized designs not only reduce LUT utilization by approximately 10\% but also entirely eliminate DSP usage. All designs met timing, while the latency-optimized designs in \texttt{hls4ml} are pipelined with more stages and have higher latency.

\begin{table*}[htbp]
  \begin{tabular}{lc|ccccccc}
    \toprule
    Strategy & Resolution  & Latency [cycles] & LUT    & DSP & FF    & $\mathrm{F}_\mathrm{max}$ [MHz] & adders \\
    \midrule
    Latency  & \trow{1.95} & 11 (67.5 ns)     & 39,413 & 522 & 6,043 & 162.9                           & (5389) \\
    DA       &             & 9 (55.4 ns)      & 37,125 & 0   & 5,547 & 162.5                           & 3193   \\
    \hline
    Latency  & \trow{2.00} & 11 (66.0 ns)     & 34,460 & 154 & 5,263 & 166.6                           & (4288) \\
    DA       &             & 9 (56.0 ns)      & 27,832 & 0   & 3,147 & 160.8                           & 2623   \\
    \hline
    Latency  & \trow{2.09} & 12 (71.4 ns)     & 24,941 & 68  & 4,677 & 168.2                           & (3336) \\
    DA       &             & 8 (47.2 ns)      & 21,778 & 0   & 3,697 & 169.5                           & 2149   \\
    \hline
    Latency  & \trow{2.20} & 13 (75.9 ns)     & 21,557 & 41  & 4,699 & 171.3                           & (2899) \\
    DA       &             & 8 (46.6 ns)      & 18,895 & 0   & 3,105 & 171.6                           & 1880   \\
    \hline
    Latency  & \trow{2.39} & 10 (57.3 ns)     & 16,918 & 27  & 2,484 & 174.5                           & (2330) \\
    DA       &             & 8 (47.0 ns)      & 14,735 & 0   & 2,215 & 170.2                           & 1574   \\
    \hline
    Latency  & \trow{2.63} & 12 (60.6 ns)     & 13,306 & 10  & 3,429 & 197.9                           & (1770) \\
    DA       &             & 8 (44.0 ns)      & 12,318 & 0   & 2,166 & 181.9                           & 1248   \\
    \bottomrule
  \end{tabular}
  \caption{Resource utilization and latency of the Muon tracking network with and without \texttt{da4ml} generated with \texttt{hls4ml}, marked with ``DA'' and ``Latency'' for strategy, respectively.
    The FPGA part is \texttt{xcvu13p-flga2577-2-e} with 160 MHz target clock frequency. Delay constraint is set to 2 for the \texttt{da4ml} optimized designs for each CMVM operation. II is 1 for all designs shown. The resolution is the truncated mean squared error (MSE) on the test set of the dataset as described in~\cite{tgc}, and the difference in resolution between different models are due to different quantization levels.
  }
  \label{tab:tgc_160MHz}
\end{table*}

\subsubsection{Particle-based Jet Tagging Network}
\begin{figure}[htbp]
  \centering
  \includegraphics[width=1.0\textwidth]{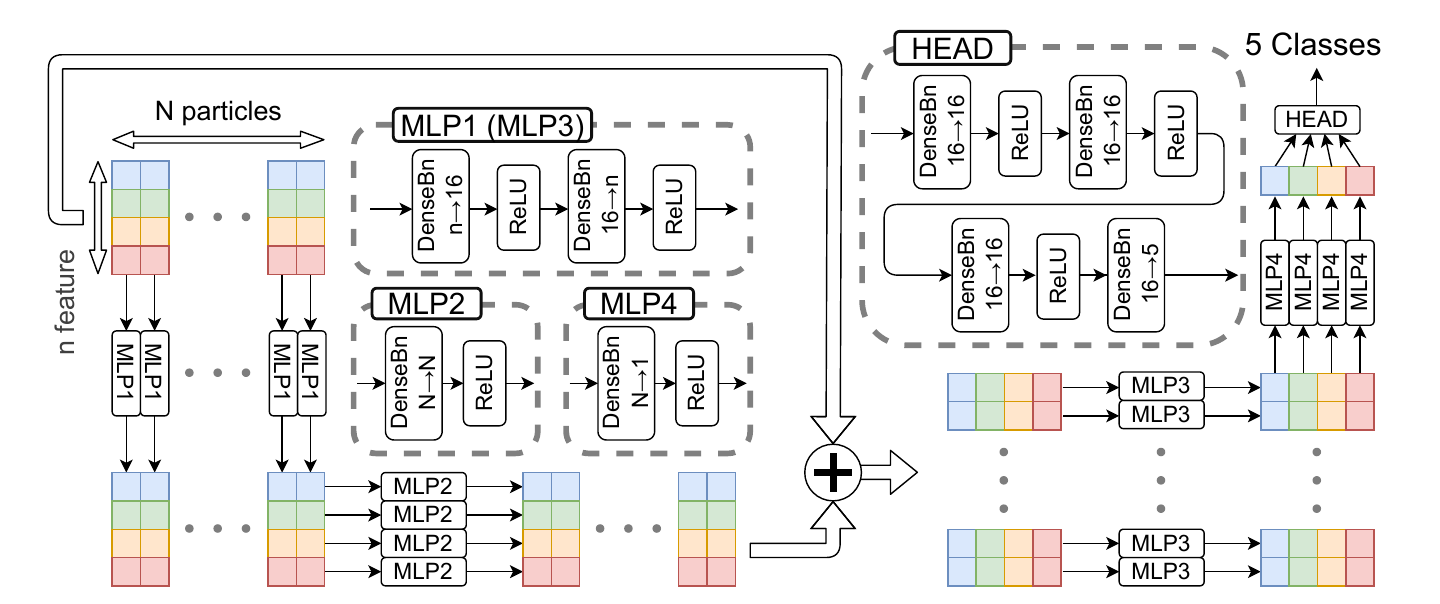}
  \caption{The architecture of the particle-based jet tagging network~\cite{mlpm-fpga}. The network is an MLP Mixer based neural network for jet tagging. The architecture is adopted from~\cite{mlpm-fpga} with the 64-particle input, and 16-features per particle ($N=64$ and $n=16$). The model consists of four MLP blocks with a single skip-connection. The implementation of each MLP and the classification head is shown in the corresponding dashed blocks. MLP1 and MLP3 act on the feature dimension; MLP2 and MLP4 act on the particle dimension. \texttt{DenseBn} represents a dense layer followed by a batch normalization layer during training, which are fused into a single layer during inference.}
  \label{mlpm}
  \Description{The architecture of an MLP-Mixer network. The diagram shows the main data path with a long skip connection arching over a series of processing blocks. Inset diagrams detail the internal structure of these blocks, showing that MLPs are applied alternatingly on the feature dimension and then on the particle dimension, which is the core concept of the mixer architecture.}
\end{figure}

The particle-based jet tagging network~\cite{mlpm-fpga} is an MLP Mixer based neural network for jet tagging. We adopt the architecture from~\cite{mlpm-fpga} with the 64-particle input, and 16-features per particle. The network is trained with the \texttt{HGQ} library~\cite{hgq}. The architecture is shown in Figure~\ref{mlpm}. For this neural network, Vitis HLS 2023.2 cannot produce designs with II of 1 clock cycle when using the original \texttt{hls4ml} implementation. We suspect the issue is related to the multidimensional linear operation implemented with Einstein Summation templates, and the usage of some pragmas may be inappropriate. However, we do not investigate this issue further as it is beyond the scope of this work.

The results are shown in Table~\ref{tab:mlpm_200MHz}. The designs are synthesized with a target clock frequency set to 200 MHz and a target II of 1. The delay constraint is set to 2 for each optimized CMVM operation. Marginal resource savings are observed for these designs for LUTs. However, the designs without \texttt{da4ml} failed to be fully pipelined with an II of 1, and the achieved $\mathrm{F}_\mathrm{max}$ is significantly lower than the target 200 MHz clock frequency.

\begin{table*}[htbp]
  \begin{tabular}{lc|ccccccc}
    \toprule
    Strategy & Accuracy      & Latency [cycles] & LUT     & DSP & FF     & $\mathrm{F}_\mathrm{max}$ [MHz] & II [cycles] & adders  \\
    \midrule
    Latency  & \trow{81.4\%} & 35 (319.6 ns)    & 137,893 & 79  & 24,785 & 109.5                           & 12          & (18536) \\
    DA       &               & 13 (62.6 ns)     & 125,694 & 0   & 26,097 & 207.6                           & 1           & 13564   \\

    \hline
    Latency  & \trow{81.0\%} & 29 (201.0 ns)    & 72,299  & 81  & 13,510 & 144.3                           & 5           & (8930)  \\
    DA       &               & 13 (63.8 ns)     & 68,157  & 0   & 13,147 & 203.9                           & 1           & 6687    \\

    \hline
    Latency  & \trow{80.3\%} & 29 (182.3 ns)    & 56,253  & 80  & 11,771 & 159.1                           & 5           & (6584)  \\
    DA       &               & 12 (56.9 ns)     & 50,657  & 0   & 11,411 & 210.9                           & 1           & 4923    \\

    \hline
    Latency  & \trow{77.3\%} & 12 (69.6 ns)     & 25,828  & 61  & 5,662  & 172.5                           & 2           & (3283)  \\
    DA       &               & 11 (53.4 ns)     & 25,002  & 0   & 5,554  & 205.8                           & 1           & 2530    \\

    \bottomrule
  \end{tabular}
  \caption{Resource utilization and latency of the particle-based jet tagging network with and without \texttt{da4ml} generated with \texttt{hls4ml}, marked with ``DA'' and ``Latency'' for strategy, respectively.
    The FPGA part is \texttt{xcvu13p-flga2577-2-e} with 200 MHz target clock frequency. Delay constraint is set to 2 for the \texttt{da4ml} optimized designs for each CMVM operation. The accuracy stands for the top-1 classification accuracy on the test set of the dataset, and the difference in accuracy between different models are due to different quantization levels.
  }
  \label{tab:mlpm_200MHz}
\end{table*}

\begin{table*}[htbp]
  \begin{tabular}{lc|ccccc}
    \toprule
    Implementation & Accuracy      & Latency [cycles] & LUT    & DSP & FF    & $\mathrm{F}_\mathrm{max}$ [MHz] \\
    \midrule
    hls4ml+DA      & \trow{76.9\%} & 4 (18.9 ns)      & 12,250 & 0   & 1,502 & 211.5                           \\
    da4ml          &               & 4 (20.0 ns)      & 12,032 & 0   & 3,113 & 200.1                           \\
    \hline
    hls4ml+DA      & \trow{76.6\%} & 3 (13.3 ns)      & 6,869  & 0   & 966   & 226.0                           \\
    da4ml          &               & 4 (19.5 ns)      & 6,153  & 0   & 1,775 & 205.4                           \\
    \hline
    hls4ml+DA      & \trow{76.5\%} & 3 (13.1 ns)      & 6,005  & 0   & 871   & 229.6                           \\
    da4ml          &               & 4 (19.8 ns)      & 5,613  & 0   & 1,539 & 202.0                           \\
    \hline
    hls4ml+DA      & \trow{76.3\%} & 3 (12.8 ns)      & 4,728  & 0   & 786   & 234.1                           \\
    da4ml          &               & 3 (14.8 ns)      & 4,129  & 0   & 903   & 202.4                           \\
    \hline
    hls4ml+DA      & \trow{76.0\%} & 3 (12.3 ns)      & 3,308  & 0   & 661   & 244.8                           \\
    da4ml          &               & 3 (14.9 ns)      & 2,814  & 0   & 628   & 201.0                           \\
    \hline
    hls4ml+DA      & \trow{75.9\%} & 3 (12.8 ns)      & 2,878  & 0   & 601   & 234.0                           \\
    da4ml          &               & 3 (14.7 ns)      & 2,485  & 0   & 683   & 204.2                           \\
    \bottomrule
  \end{tabular}
  \caption{
    Resource utilization and latency of the high-level feature jet tagging network generated with \texttt{hls4ml} with \texttt{da4ml}, and with \texttt{da4ml} directly via Verilog, marked with ``hls4ml+DA'' and ``da4ml'', respectively.
    The FPGA part is \texttt{xcvu13p-flga2577-2-e} with 200 MHz target clock frequency. Delay constraint is set to 2 for the \texttt{da4ml} optimized designs for each CMVM operation. The Verilog designs are pipelined every 5 adders. II is 1 for all designs shown.
  }
  \label{tab:v_jsc_200MHz}
\end{table*}

\begin{table*}[htbp]
  \begin{tabular}{lc|ccccc}
    \toprule
    Implementation & Accuracy      & Latency [cycles] & LUT    & DSP & FF     & $\mathrm{F}_\mathrm{max}$ [MHz] \\
    \midrule
    hls4ml+DA      & \trow{76.9\%} & 31 (44.1 ns)     & 12,682 & 0   & 19,056 & 702.2                           \\
    da4ml          &               & 20 (34.0 ns)     & 12,298 & 0   & 13,664 & 588.6                           \\
    \hline
    hls4ml+DA      & \trow{76.6\%} & 26 (37.4 ns)     & 7,392  & 0   & 11,462 & 695.9                           \\
    da4ml          &               & 16 (27.5 ns)     & 6,944  & 0   & 7,419  & 582.4                           \\
    \hline
    hls4ml+DA      & \trow{76.5\%} & 25 (35.4 ns)     & 6,448  & 0   & 10,109 & 707.2                           \\
    da4ml          &               & 17 (23.1 ns)     & 6,165  & 0   & 7,207  & 735.8                           \\
    \hline
    hls4ml+DA      & \trow{76.3\%} & 22 (30.2 ns)     & 5,019  & 0   & 7,682  & 729.4                           \\
    da4ml          &               & 15 (20.8 ns)     & 4,655  & 0   & 5,258  & 722.0                           \\
    \hline
    hls4ml+DA      & \trow{76.0\%} & 23 (32.6 ns)     & 3,602  & 0   & 5,693  & 706.2                           \\
    da4ml          &               & 14 (20.6 ns)     & 3,127  & 0   & 3,587  & 681.2                           \\
    \hline
    hls4ml+DA      & \trow{75.9\%} & 21 (24.9 ns)     & 2,908  & 0   & 4,720  & 844.6                           \\
    da4ml          &               & 13 (19.0 ns)     & 2,734  & 0   & 3,139  & 685.4                           \\
    \bottomrule
  \end{tabular}
  \caption{
    Resource utilization and latency of the high-level feature jet tagging network generated with \texttt{hls4ml} with \texttt{da4ml}, and with \texttt{da4ml} directly via Verilog, marked with ``hls4ml+DA'' and ``da4ml'', respectively.
    The FPGA part is \texttt{xcvu13p-flga2577-2-e} with 1 GHz target clock frequency. Delay constraint is set to 2 for the \texttt{da4ml} optimized designs for each CMVM operation. The Verilog designs are pipelined every adder. II is 1 for all designs shown.
  }
  \label{tab:v_jsc_1GHz}
\end{table*}

\begin{table*}[htbp]
  \begin{tabular}{lc|ccccc}
    \toprule
    Implementation & Accuracy      & Latency [cycles] & LUT     & DSP & FF     & $\mathrm{F}_\mathrm{max}$ [MHz] \\
    \midrule

    hls4ml+DA      & \trow{81.4\%} & 13 (62.6 ns)     & 125,694 & 0   & 26,097 & 207.6                           \\
    da4ml          &               & 11 (54.8 ns)     & 120,512 & 0   & 28,284 & 200.8                           \\
    \hline

    hls4ml+DA      & \trow{81.0\%} & 13 (63.8 ns)     & 68,157  & 0   & 13,147 & 203.9                           \\
    da4ml          &               & 11 (54.8 ns)     & 61,771  & 0   & 16,031 & 200.6                           \\
    \hline

    hls4ml+DA      & \trow{80.3\%} & 12 (56.9 ns)     & 50,657  & 0   & 11,411 & 210.9                           \\
    da4ml          &               & 10 (48.4 ns)     & 47,778  & 0   & 12,515 & 206.8                           \\
    \hline

    hls4ml+DA      & \trow{77.3\%} & 11 (53.4 ns)     & 25,002  & 0   & 5,554  & 205.8                           \\
    da4ml          &               & 9 (41.9 ns)      & 24,285  & 0   & 6,262  & 214.8                           \\
    \bottomrule
  \end{tabular}
  \caption{
    Resource utilization and latency of the particle-based jet tagging network generated with \texttt{hls4ml} with \texttt{da4ml}, and with \texttt{da4ml} directly via Verilog, marked with ``hls4ml+DA'' and ``da4ml'', respectively.
    The FPGA part is \texttt{xcvu13p-flga2577-2-e} with 200 MHz target clock frequency. Delay constraint is set to 2 for the \texttt{da4ml} for each CMVM operation. The Verilog designs are pipelined every 5 adders. II is 1 for all designs shown. The accuracy stands for the top-1 classification accuracy on the test set of the dataset, and the difference in accuracy between different models are due to different quantization levels.
  }

  \label{tab:v_mlpm_200MHz}

\end{table*}

\subsection{RTL Generation Performance}

To evaluate the performance of the standalone RTL code generation in \texttt{da4ml}, we show the performance of the high-level feature jet tagging network with 200 MHz and 1 GHz target clock frequencies in Table~\ref{tab:v_jsc_200MHz} and Table~\ref{tab:v_jsc_1GHz}, and the results for the particle-based jet tagging network with MLP Mixer in Table~\ref{tab:v_mlpm_200MHz}, respectively. For RTL generation, we use Verilog for benchmark in this section. The designs generated with \texttt{hls4ml} are marked with ``hls4ml+DA'', and the designs generated directly by \texttt{da4ml} with Verilog are marked with ``da4ml''. All designs are pipelined with an II of 1, and the delay constraint is set to 2 for each CMVM operation. For the standalone generated RTL designs, we pipeline every 5 adders for the 200 MHz target clock frequency and every adder for the 1 GHz target clock frequency. Global retiming is enabled in Vivado for the Verilog designs.

The results show that the designs generated in RTL directly with \texttt{da4ml} can achieve lower LUT consumption by $5-20$\% compared to the designs generated in conjunction with \texttt{hls4ml} and synthesized with Vitis HLS. However, the achieved $\mathrm{F}_\mathrm{max}$ is generally slightly lower. In particular, it can be noticed that the designs generated in conjunction with \texttt{hls4ml} are pipelined with more stages than the adder graph depth, which suggests that the HLS compiler is further breaking down the graph for timing optimization, a process not performed in the standalone code generation. We suspect this is the reason for the lower $\mathrm{F}_\mathrm{max}$ in the designs generated with \texttt{da4ml}.

It is also worth noting that generating designs with \texttt{da4ml} directly in RTL can significantly reduce compilation time: for designs with a large amount of code, HLS can take a very long time. For instance, the synthesis time for the particle-based jet tagging networks\footnote{All-inclusive time, until routing finishes} with Vitis HLS and Vivado 2023.2 is around 17 hours without memory bottleneck when optimized with \texttt{da4ml}. However, the designs of the same networks generated directly in Verilog with \texttt{da4ml} only take around 26 minutes to synthesize with Vivado 2023.2. The peak memory usage also significantly reduced from $\sim 70$ GB to $\sim 10$ GB. These results suggest that \texttt{da4ml}'s standalone code generation can facilitate fast prototyping and allow for faster iterations in software-hardware co-design.

\begin{table*}[htbp]
  \begin{adjustbox}{width=1.0\textwidth,center=\textwidth}
    \begin{tabular}{l|ccccccc}
      \multicolumn{8}{c}{\textbf{High-level feature jet tagging network}}                                                                                                                  \\
      \midrule
      Implementation                                                & Accuracy          & Latency [cycles]   & LUT       & DSP   & FF      & $\mathrm{F}_\mathrm{max}$ [MHz] & II [cycles] \\
      \midrule
      \texttt{\textbf{HGQ+da4ml (HLS)}}                             & 76.9\%            & 31 (44.1 ns)       & 12,682    & 0     & 19,056  & 702.2                           & 1           \\
      \texttt{\textbf{HGQ+da4ml (RTL)}}                             & 76.5\%            & 17 (23.1 ns)       & 6,165     & 0     & 7,207   & 735.8                           & 1           \\
      \texttt{HGQ+hls4ml}                                           & 76.9\%            & 42 (57.6 ns)       & 16,081    & 57    & 26,484  & 729.4                           & 1           \\
      \texttt{HGQ+hls4ml}                                           & 76.5\%            & 35 (67.2 ns)       & 8,548     & 30    & 14,418  & 520.8                           & 1           \\
      \texttt{QKeras+hls4ml} [ICFPT'23]~\cite{dsp-prune}$^{***}$    & 76.3\%            & 15 (105 ns)        & 5,504     & 175   & 3,036   & 142.9                           & 2           \\
      \texttt{DWN} [ICLR'24]~\cite{dwn}$^{**}$                      & 76.3\%            & 10 (14.4 ns)       & 6,302     & 0     & 4,128   & 695.                            & 1           \\
      \texttt{MetaML-Pro} [Arxiv'25]~\cite{metamlpro}$^{*}$         & 76.1\%            & 10 (50 ns)         & 13,042    & 70    & N/A     & 200                             & 1           \\
      \texttt{NeuraLUT-Assemble} [FCCM'25]~\cite{neuralut-assemble} & 76.0\%            & 2 (2.1 ns)         & 1,780     & 0     & 540     & 940.                            & 1           \\
      \texttt{TreeLUT} [FPGA'25]~\cite{treelut}                     & 75.6\%            & 2 (2.7 ns)         & 2,234     & 0     & 347     & 735.                            & 1           \\
      \midrule
      \multicolumn{8}{c}{\textbf{High-level feature jet tagging network (CERN Box)}}                                                                                                       \\
      \midrule
      Implementation                                                & Accuracy          & Latency [cycles]   & LUT       & DSP   & FF      & $\mathrm{F}_\mathrm{max}$ [MHz] & II [cycles] \\
      \midrule
      \texttt{\textbf{HGQ+da4ml (RTL)}}                             & 75.2\%            & 37.8               & 8,703     & 0     & 10,008  & 503.0                           & 1           \\
      \texttt{\textbf{HGQ+da4ml (RTL)}}                             & 75.0\%            & 27.4               & 5,636     & 0     & 6,218   & 656.2                           & 1           \\
      \texttt{NeuraLUT-Assemble} [FCCM'25]~\cite{neuralut-assemble} & 75.0\%            & 5.7                & 8,539     & 0     & 1,332   & 352.                            & 1           \\
      \texttt{NeuraLUT-Assemble} [FCCM'25]~\cite{neuralut-assemble} & 75.0\%            & 7.0                & 8,535     & 0     & 2,717   & 994.                            & 1           \\
      \texttt{AmigoLUT-NeuraLUT} [FPGA'25]~\cite{amigolut}          & 74.4\%            & 9.6                & 42,742    & 0     & 4,717   & 520.                            & 1           \\
      \texttt{PolyLUT-Add} [FPL'24]~\cite{polylut-add}              & 75.\%             & 16                 & 36,484    & 0     & 1,209   & 315.                            & 1           \\
      \texttt{NeuraLUT} [FPL'24]~\cite{neuralut}                    & 75.\%             & 14                 & 92,357    & 0     & 4,885   & 368.                            & 1           \\
      \texttt{PolyLUT} [TC'25]~\cite{polylut}                       & 75.1\%            & 25                 & 246,071   & 0     & 12,384  & 203.                            & 1           \\
      \texttt{LogicNets} [FPL'20]~\cite{logicnets}                  & 72.\%             & 13                 & 37,931    & 0     & 810     & 427.                            & 1           \\
      \midrule
      \multicolumn{8}{c}{\textbf{Muon tracking network}}                                                                                                                                   \\
      \midrule
      Implementation                                                & Resolution [mrad] & Latency [cycles]   & LUT       & DSP   & FF      & $\mathrm{F}_\mathrm{max}$ [MHz] & II [cycles] \\
      \midrule
      \texttt{\textbf{HGQ+da4ml (HLS)}}                             & 1.95              & 9 (55.4 ns)        & 37,125    & 0     & 5,547   & 162.5                           & 1           \\
      \texttt{HGQ+hls4ml}~\cite{hgq}                                & 1.95              & 11 (67.5 ns)       & 39,413    & 522   & 6,043   & 162.9                           & 1           \\
      \texttt{QKeras+hls4ml} [NIMA'23]~\cite{tgc}$^{***}$           & 1.95              & 17 (106.3 ns)      & 37,867    & 1,762 & 8,443   & 160                             & 1           \\
      \midrule
      \multicolumn{8}{c}{\textbf{SVHN classification network}}                                                                                                                             \\
      \midrule
      Implementation                                                & Accuracy          & Latency [cycles]   & LUT       & DSP   & FF      & $\mathrm{F}_\mathrm{max}$ [MHz] & II [cycles] \\
      \midrule
      \texttt{\textbf{HGQ+da4ml (HLS)}}                             & 93.9\%            & 1,045 (5,159.2 ns) & 53,425    & 0     & 20,048  & 202.6                           & 1,029       \\
      \texttt{HGQ+hls4ml}~\cite{hgq}                                & 93.9\%            & 1,050 (5,058.9 ns) & 69,407    & 58    & 27,853  & 207.6                           & 1,029       \\
      \texttt{QKeras+hls4ml} [MLST'21]~\cite{fast_cnn}$^{*}$        & 94.\%             & 1,035 (5,175 ns)   & 111,152   & 174   & 32,554  & 200                             & 1,030       \\
      \texttt{QKeras+hls4ml} [ICFPT'23]~\cite{dsp-prune}$^{***}$    & 92.4\%            & 5,447 (43,576 ns)  & 59,279    & 1,215 & 46,584  & 125                             & N/A         \\
      \midrule
      \multicolumn{8}{c}{\textbf{Particle-based jet tagging network}}                                                                                                                      \\
      \midrule
      Implementation                                                & Accuracy          & Latency [cycles]   & LUT       & DSP   & FF      & $\mathrm{F}_\mathrm{max}$ [MHz] & II [cycles] \\
      \midrule
      \texttt{\textbf{HGQ+da4ml (RTL)}}                             & 81.4\%            & 11 (54.8 ns)       & 120,512   & 0     & 28,284  & 200.8                           & 1           \\
      \texttt{LL-GNN} [TEC'23]~\cite{jedi-fpga}$^{*}$               & 81.2\%            & 181 (905 ns)       & 815k      & 8,986 & 189k    & 200                             & 150         \\
      \texttt{QKeras+hls4ml, DS} [MLST'24]~\cite{ds-fpga}$^{*}$     & $\le 75.9$\%      & 26 (130 ns)        & 903,284   & 434   & 358,754 & 200                             & 2           \\
      \texttt{QKeras+hls4ml, GNN} [MLST'24]~\cite{ds-fpga}$^{*}$    & $\le 75.8$\%      & 32 (160 ns)        & 1,162,104 & 2,120 & 761,061 & 200                             & 3           \\
      \midrule
    \end{tabular}
  \end{adjustbox}
  \caption{
    Resource and performance of various neural network tasks with different implementations. When the neural network is quantized with HGQ~\cite{hgq}, we show that traditional neural networks may achieve competitive resource utilization, throughput, and superior accuracy compared with the methods designed specifically for FPGAs. All works are using either \texttt{xcvu13p-flga2577-2-e} or \texttt{xcvu9p-flgb2104-2-i}, where the results are comparable. \texttt{da4ml (HLS)} refers to using \texttt{da4ml} with the \texttt{hls4ml} workflow, while \texttt{da4ml (RTL)} refers to the standalone workflow. Works marked with $^{*}$ report the results after logic synthesis without place and route. We put the \textit{target clock frequency} for $\mathrm{F}_\mathrm{max}$ and used it to compute the latency for these works. DWN results$^{**}$ are cited from~\cite{neuralut-assemble} instead, as the original paper omits preprocessing steps for the FPGA hardware implementation, which are added in~\cite{neuralut-assemble} for a fair comparison. Works marked with $^{***}$ reports resource utilization after place and routes, but did not report $\mathrm{F}_\mathrm{max}$. We use the achieved target clock as $\mathrm{F}_\mathrm{max}$ for these works.
  }
  \label{tab:compare}

\end{table*}

\subsection{Comparison to Other Methods}
While this work focuses on distributed arithmetic optimization for CMVM operations in FPGA-based neural networks, alternative methods exist for implementing machine learning-based models on FPGAs. In this section, we briefly compare neural networks optimized with \texttt{da4ml} to other methods. We show the comparison in Table~\ref{tab:compare}, where we compare the performance of the networks shown in earlier sections against other state-of-the-art methods. To make a more holistic comparison, we include another variant of the high-level feature jet tagging task, marked as ``CERN Box'', which has different pre-processing and has more noisy inputs compared to the one shown in earlier sections~\cite{neuralut-assemble}.

Comparing to other conventional neural network based methods shown in Table~\ref{tab:compare}, our proposed \texttt{da4ml} framework demonstrates significant advantage across diverse workloads, offering a favorable balance between accuracy, latency, and hardware efficiency. Notably, our designs consistently eliminate DSP usage while maintaining competitive or superior accuracy, making them suited for resource-constrained FPGA designs. Similarly, in the muon tracking and SVHN classification tasks, our method matches the accuracy of existing approaches while reducing LUT and flip-flop utilization. Furthermore, the particle-based jet tagging design achieves the highest accuracy (81.4\%) while consuming over 10$\times$ fewer LUTs and FFs than state-of-the-art GNN-based implementations. Across all tasks, our designs maintain a consistent II of 1 cycle, ensuring high-throughput operation. These results underscore the generalizability and practicality of our approach for low-latency, resource-efficient FPGA deployment in both traditional and physics-inspired neural networks.

LUT-based methods have been proposed as an alternative to traditional neural networks for FPGA implementations in the latency- and resource-constrained scenarios. Many works cited in the table are LUT-based methods, including TreeLUT~\cite{treelut}, NeuraLUT-Assemble~\cite{neuralut-assemble}, DWN~\cite{dwn}, MetaML-Pro~\cite{metamlpro}, AmigoLUT-NeuraLUT~\cite{amigolut}, PolyLUT-Add~\cite{polylut-add}, PolyLUT~\cite{polylut}, and LogicNets~\cite{logicnets}. While these methods are proposed as more resource-efficient alternatives, it is important to note that they also come with trade-offs in terms of accuracy and generalizability. As of the timing of writing, these methods have only been demonstrated on small-scale problems requiring $\sim 1$ k LUTs, and their applicability to more complex problems requiring $\sim 100$ k LUTs is not demonstrated by any existing work.

While some LUT-based methods can achieve exceptional efficiency in terms of resource (LUTs) utilization and throughput, they do not necessarily outperform traditional neural networks in accuracy-resource trade-offs. For instance, while NeuraLUT-Assemble~\cite{neuralut-assemble} can achieve better resource-accuracy on the simpler high-level feature jet tagging task, it suffers from significant accuracy degradation compared to traditional neural networks optimized with \texttt{da4ml}. Here, for the JSC OpenML and CERNBox datasets, the dataset constructed for benchmarking purposes, and the obtained accuracy is clustered around 72-77\% or 72-76\%, respectively, for a large variety of different methods. We also observe that the resource consumption changes rapidly with small accuracy changes in this range. Hence, for these datasets, we consider any accuracy difference with statistical significance, i.e., $> 0.1\%$, to be significant. On the CERN Box variant, NeuraLUT-Assemble's resource-accuracy trade-off is also outperformed by traditional neural networks optimized with \texttt{da4ml}. Other LUT-based methods, such as TreeLUT~\cite{treelut}, AmigoLUT-NeuraLUT~\cite{amigolut}, PolyLUT-Add~\cite{polylut-add}, PolyLUT~\cite{polylut}, and LogicNets~\cite{logicnets}, are significantly outperformed by these two methods in significant margins.

Importantly, as traditional neural networks have already been deployed in production systems at the LHC~\cite{axol1tl, cicada}, while LUT-based methods are still in the research phase, we believe that traditional neural networks optimized with \texttt{da4ml} provide a more practical solution for real-world applications at the current stage.

\section{Conclusion}
\label{sec:conclusion}

This work presents an efficient algorithm for optimizing CMVM operations with distributed arithmetic for FPGA-based neural networks. We further present the \texttt{da4ml} library, which implements the algorithm efficiently and tightly integrates it with the \texttt{hls4ml} library for easy adoption. We then demonstrate the performance of the algorithm on a set of neural networks, including dense, convolutional, and MLP-Mixer-based networks. The results show that designs optimized with \texttt{da4ml} can simultaneously reduce resource utilization and timing on realistic, heavily quantized, and sparse neural networks for physics applications. We also show that for a limited set of networks, \texttt{da4ml} can generate Verilog designs directly for fast prototyping and software-hardware co-design. The results show that designs generated directly with \texttt{da4ml} could further reduce resource utilization and synthesis time significantly at the expense of a slightly lower $\mathrm{F}_\mathrm{max}$. We hope this work will facilitate the design of efficient neural networks for FPGAs, especially in the context of physics applications, and provide a solid foundation for future work on hardware-aware neural network design. Future work includes extending the set of supported NN operators, such as RNNs and Transformers, and raising the RTL workflow's frontend from manual functional specs to ONNX/Torch import, and integrating \texttt{da4ml} with the open-source HLS tool \texttt{Bambu}~\cite{bambu} to further broaden applicability, and validating on additional physics triggers as well as non-HEP domains (astrophysics, finance, healthcare) across a wider range of devices.

\section{Acknowledgments}

Chang Sun and Maria Spiropulu are partially supported by the U.S. Department of Energy (DOE), Office of Science, Office of High Energy Physics grant DE-SC0011925. Chang Sun is partially supported by the NSF ACCESS Grant number PHY240298. Zhiqiang Que and Wayne Luk are supported by the United Kingdom EPSRC (grant numbers UKRI256, EP/V028251/1, EP/N031768/1, EP/S030069/1, and EP/X036006/1). Vladimir Loncar is supported by the Eric \& Wendy Schmidt Fund for Strategic Innovation through the CERN Next Generation Triggers project under grant agreement number SIF-2023-004.

\bibliographystyle{ACM-Reference-Format}
\bibliography{bibfile}

\end{document}